\documentclass[twocolumn,aps,prb]{revtex4-1} 
\usepackage[english]{babel}
\usepackage{amsmath}
\usepackage{amsfonts}
\usepackage{hyperref}
\usepackage{mathtools}
\usepackage[utf8]{inputenc} 
\usepackage{graphicx}
\usepackage{xfrac}
\usepackage{float} 
\usepackage{siunitx} 
\usepackage{natbib}

\usepackage{comment}

\hypersetup{colorlinks,
            linkcolor=blue,
            citecolor=blue,
            urlcolor=black
           }

\DeclareSIUnit{\rydberg}{Ry}
\DeclareSIUnit{\bohr}{\ensuremath{a_0}}

\newcommand{\bk}{\ensuremath{\mathbf{k}}}       
\newcommand{\pbk}{\ensuremath{^{\mathbf{k}}}}   
\newcommand{\bq}{\ensuremath{\mathbf{q}}}       
\newcommand{\br}{\ensuremath{\mathbf{r}}}       
\newcommand{\bR}{\ensuremath{\mathbf{R}}}       
\newcommand{\bT}{\ensuremath{\mathbf{T}}}       
\newcommand{\bG}{\ensuremath{\mathbf{G}}}       
\newcommand{\bZ}{\mathbf{0}}                    
\newcommand{\conj}[1]{\overline{#1}}            
\newcommand{\kmt}[1]{#1$\times$#1$\times$#1}    

\providecommand{\abs}[1]{\lvert#1\rvert}

\DeclarePairedDelimiter\ket{\lvert}{\rangle}
\DeclarePairedDelimiter\evo{\langle}{\rangle}
\DeclarePairedDelimiterX\inn[2]{\langle}{\rangle}{#1 \delimsize| #2}
\DeclarePairedDelimiterX\outt[2]{\lvert}{\rvert}{#1 \delimsize\rangle\!\delimsize\langle #2}
\DeclarePairedDelimiterX\outo[1]{\lvert}{\rvert}{#1 \delimsize\rangle\!\delimsize\langle #1}
\DeclarePairedDelimiterX\evt[2]{\langle}{\rangle}{#2 \delimsize\lvert #1 \delimsize\rvert #2}
\DeclarePairedDelimiterX\mel[3]{\langle}{\rangle}{#1 \delimsize\lvert #2 \delimsize\rvert #3}

\DeclareMathOperator{\erfc}{erfc}
\DeclareMathOperator{\erf}{erf}
\DeclareMathOperator{\tr}{tr}

\let\Re\relax

\newcommand{\Re}{\ensuremath{\mathrm{Re}}}

\DeclareUnicodeCharacter{2010}{-}
\DeclareUnicodeCharacter{2009}{ }

\begin{document}

\title{Localized orbital scaling correction for periodic systems}
\author{Aaron Mahler}
\affiliation{Duke University, Department of Physics, Durham, NC 27708}
\author{Jacob Williams} 
\affiliation{Duke University, Department of Chemistry, Durham, NC 27708}
\author{Neil Qiang Su} 
\affiliation{Department of Chemistry, Key Laboratory of Advanced Energy Materials Chemistry (Ministry of Education) and
Renewable Energy Conversion and Storage Center (RECAST), Nankai University, Tianjin 300071, China}
\affiliation{Duke University, Department of Chemistry, Durham, NC 27708}
\author{Weitao Yang} 
\email{weitao.yang@duke.edu}
\affiliation{Duke University, Department of Chemistry, Durham, NC 27708}
\affiliation{Duke University, Department of Physics, Durham, NC 27708}
\date{\today}

\begin{abstract}
Density functional theory offers accurate structure prediction at acceptable
computational cost,  but commonly used approximations suffer from delocalization
error; this results in inaccurate predictions of quantities such as energy band
gaps of finite and bulk systems, energy level alignments, and electron
distributions at interfaces. The localized orbital scaling correction (LOSC) was
developed to correct delocalization error by using orbitals localized in space
and energy. These localized orbitals span both the occupied and unoccupied
spaces and can have fractional occupations in order to correct both the total
energy and the one-electron energy eigenvalues. We extend the LOSC method to
periodic systems, in which the localized orbitals employed are dually localized
Wannier functions. In light of the effect of the bulk environment on the
electrostatic interaction between localized orbitals, we modify the LOSC energy
correction to include a screened Coulomb kernel. For a test set of
semiconductors and large-gap insulators, we show that the screened LOSC
(sLOSC) method consistently improves the band gap compared to the parent density
functional approximation. 
\end{abstract}

\maketitle

\section{Introduction}
The cost of solving the electronic Schr\"odinger equation scales exponentially
with the size of the system, exceeding the computational resources available on
the planet for any system larger than a few tens of electrons.
\cite{kohn_nobel_1999} Density functional theory (DFT) sidesteps this
exponential cost by treating the electron density as the fundamental variable 
instead of computing the wavefunction directly and by constructing an auxiliary
noninteracting reference system sharing the density of the physical system.
\cite{hohenberg_inhomogeneous_1964,kohn_self-consistent_1965} Due to the
accuracy attainable at a cost only cubic in the number of electrons $N$, DFT has
become a mainstay of computational chemistry and materials science.
\cite{becke_densityfunctional_1993,lee_development_1988,perdew_generalized_1996,
      van_noorden_top_2014}
While DFT is exact in theory, the form of the universal exchange-correlation
functional is unknown, and density functional approximations (DFAs) must be used
in practice. Commonly used DFAs suffer from systematic delocalization and static
correlation errors. \cite{cohen_insights_2008,cohen_challenges_2012} The
delocalization error underlies the failure of DFAs to describe energy band gaps
of finite and bulk systems, energy level alignments, and electron distributions
at interfaces. \cite{mori-sanchez_localization_2008} Overcoming delocalization
error remains an active and challenging research effort.

Connecting single-particle orbital energies $\epsilon$ to observable quantities
was another longstanding question in Kohn--Sham DFT. As a contrast,
\citet{koopmans_uber_1934} showed in 1934 that the Hartree--Fock ionization
potential (IP) and electron affinity (EA) are given under the frozen orbital
aproximation by the negative of the highest occupied and lowest unoccupied
molecular orbital eigenvalues respectively. A series of three results
established a rigorous connection for DFT.

First, \citet{janak_proof_1978} derived a link between the Kohn--Sham orbital
energies $\epsilon_m$ and the total energy $E$, viewed as a function of the
orbital occupation numbers $n_m$:
\begin{equation}
    \epsilon_m = \frac{\partial E}{\partial n_m}.
\end{equation}
However, $\partial E/\partial n_m$ was not yet linked to a physical observable.

A few years later, Perdew, Parr, Levy, and Balduz
\cite{perdew_density-functional_1982} showed that $E$ is piecewise linear in
the number of electrons $N$ when computed with the exact functional; that is,
for all $\abs{\delta} \leq 1$, we have
\begin{equation}
    E(N+\delta) = 
        \begin{cases} 
            (1+\delta)E(N) - \delta E(N-1) & \delta < 0, \\
            (1-\delta)E(N) + \delta E(N+1) & \delta \geq 0.
        \end{cases}
\end{equation}
This relationship, called the PPLB condition, connects the chemical potential
$\mu(N) = \partial E/\partial N$ to the IP and EA; observe that
\begin{equation} \label{eqn:chemPot}
\mu(N)=
\begin{cases} 
      -I(N) = E(N) - E(N-1) & \partial N<0, \\
      -A(N) = E(N+1) - E(N) & \partial N>0.
\end{cases}
\end{equation}

Finally, \citet{cohen_fractional_2008} proved that the chemical potential is
given by the partial derivative of $E$ with respect to the frontier orbital
eigenvalues,
\begin{equation}
\mu(N)= \frac{\partial E}{\partial n_f}.
\end{equation}
Crucially, $f$ labels not only the highest unoccupied molecular orbital (HOMO)
if $\partial N < 0$, but also the lowest unoccupied molecular orbital (LUMO)
if $\partial N > 0$; this was the first time a physical meaning for the
energy of the Kohn--Sham LUMO was established. This result holds for any local
functional continuous in the electron density, as well as any nonlocal functional
continuous in the Kohn--Sham density matrix; in the latter case, the work also
extends Janak's theorem to the eigenvalues from the generalized Kohn--Sham
equations.

Combining these three results, we see that
\begin{equation} \label{eqn:chemPot2janak}
    \mu(N) = 
    \begin{cases}
        -I(N) = \epsilon_{\tiny{\text{HOMO}}} & \partial N < 0, \\
        -A(N) = \epsilon_{\tiny{\text{LUMO}}} & \partial N \geq 0.
    \end{cases}
\end{equation}
Thus, the frontier eigenvalues obtained from an $N$-electron DFT calculation
correspond rigorously to physically relevant quantities;\cite{cohen_fractional_2008}
if the PPLB condition is obeyed and the functional predicts the exact energies for
$N-1$, $N$, and $N+1$ electrons, the correspondence is exact.

A feature derivable from these quantities is the fundamental or integer gap,
defined as the difference between the IP and the EA:
\begin{equation} \label{eqn:intgap}
\begin{split}
    E_{\text{gap}}^{\text{integer}} &= I - A \\
    &= E(N-1) - 2E(N) + E(N+1).
\end{split}
\end{equation}
$E_{\text{gap}}^{\text{integer}}$ quantifies the difference between positively
and negatively ionizing the system and is a crucial part of the accurate
modeling of semiconductor electronic structure. If the PPLB condition is 
obeyed, Eqs.\ (\ref{eqn:chemPot}) and (\ref{eqn:chemPot2janak}) also allow the
gap to be computed from a single $N$-electron calculation as the discontinuity
in the chemical potential; in this form, it is called the derivative gap,
defined as
\begin{equation} \label{eqn:derGap}
\begin{split} 
E^{\text{deriv}}_{\text{gap}} &= 
\left.\frac{\partial E}{\partial N}\right\rvert_+ - 
\left.\frac{\partial E}{\partial N}\right\rvert_- \\
&= \epsilon_{\text{\tiny{LUMO}}} - \epsilon_{\text{\tiny{HOMO}}}.
\end{split}
\end{equation} 
If the PPLB condition is obeyed, the derivative gap and the integer gap are 
equal.\cite{cohen_fractional_2008} In bulk systems with periodic boundary
conditions, the PPLB condition is satisfied by any DFA
continuous in the Kohn--Sham density or density matrix, regardless of systematic
errors in its definition; \cite{mori-sanchez_localization_2008} thus, the
fundamental gap of bulk systems can be predicted by the (generalized) Kohn--Sham
orbital gap, for functionals continuous in the density (density matrix), as in
Eq.\ (\ref{eqn:derGap}). In finite systems, however, the PPLB condition is not in
general obeyed, and the gap computed from Eq.\ (\ref{eqn:derGap}) may differ from
that computed by calculating the $(N \pm 1)$-electron energies to obtain the
integer gap as in Eq.\ (\ref{eqn:intgap}), the $\Delta$SCF method.

\subsection{Delocalization error}
The delocalization error has a dramatic size-dependent manifestation. In finite
systems, standard DFAs fail to obey the PPLB linearity condition, so the
derivative gap is not equal to the integer gap. This is due to the error in the
approximate exchange-correlation functional, which nearly always yields $E$
convex in $N$, underestimating the piecewise linearity prescribed by the PPLB
condition. This convex deviation has been identified as the cause for an
unphysical smearing of the electron density in space, as well as underestimation
of the total energy in a delocalized electron density; thus, we may identify it
with delocalization error, as exhibited in small systems. In bulk systems, the
delocalized nature of the orbitals produces a total energy linear with respect
to fractional charge, yielding no deviation from the PPLB condition; however,
delocalization error manifests as an incorrect slope of the $E(N)$ line at
integer $N$. \cite{mori-sanchez_localization_2008}

The effects of delocalization error include the underestimation of band gaps and
reaction barriers, \cite{merkle_singly_1992} undervaluation of dissociation
curves, \cite{zhang_challenge_1998,mori-sanchez_many-electron_2006,
             perdew_exchange_2007} overestimation of conductance and
polarizability, \cite{korzdorfer_self-interaction_2008} and incorrect energy
level alignment and charge transfer across interfaces. 
\cite{jones_density_1989,
      becke_perspective_2014} To capture the full derivative discontinuity and
hence the band gap, it has been shown that the exact functional, whether local or
nonlocal, cannot be a differentiable functional of the electron density or of the
Kohn-Sham density matrix. \cite{mori-sanchez_discontinuous_2009,
                               yang_derivative_2012}
To reduce the systematic delocalization error, many approaches have been
developed, including range-separated functionals,
\cite{savin_density_1995,savin_degeneracy_1996,iikura_long-range_2001,
      yanai_new_2004,vydrov_assessment_2006,chai_long-range_2008,
      baer_tuned_2010}
the screened range-separated hybrid functional, \cite{wing_band_2021}
self-interaction error corrected functionals,
\cite{perdew_self-interaction_1981,mori-sanchez_many-electron_2006,
      mori-sanchez_self-interaction-free_2006,perdew_density_2008,
      schmidt_one-electron_2014,pederson_communication_2014,
      schmidt_one-_2016,yang_full_2017}
Koopmans-compliant functionals,
\cite{borghi_koopmans-compliant_2014,colonna_koopmans-compliant_2019}
and generalized transition state methods, \cite{anisimov_transition_2005} along
with related developments using localized Wannier functions.
\cite{ma_using_2016}

The localized orbital scaling correction (LOSC) method was developed to
eliminate delocalization error systematically.
\cite{li_localized_2018,su_preserving_2020} Previous incarnations of LOSC were
implemented for molecular systems with real orbitals and the boundary condition
$\lim_{|\br|\to \infty} \rho(\br) = 0$. They accurately model IP, EA,
photoemission spectra, dissociation curves, and polarizabilities,
as well as restore size-consistency. 
\cite{li_localized_2018,su_preserving_2020,mei_self-consistent_2020,
      mei_describing_2021}
In this work, we extend LOSC to periodic boundary conditions (PBCs) and complex
orbitals. Additionally, we introduce a screened Coulomb interaction to the LOSC
energy correction to enable the accurate computation of bulk system band
structures.

\subsection{Periodic boundary conditions}
In PBCs, the eigenfunctions of the single-particle Hamiltonian are known as
Bloch orbitals; they satisfy
$h_{\text{s}} \ket{\psi_n\pbk} = \epsilon_n\pbk \ket{\psi_n\pbk}$. The
Bloch orbitals are also eigenfunctions of the unit cell translation operator,
so they take the form $\ket{\psi_n\pbk} = e^{i\bk\cdot\br} \ket{u_n\pbk}$,
where $\ket{u_n\pbk}$ has the periodicity of the unit cell and $\bk$ is a
point in the Brillouin zone (the reciprocal-space unit cell).
\cite{bloch_uber_1929,ashcroft_solid_2003} The Bloch orbitals obey the
normalization convention
$\inn{\psi_m^{\bq}}{\psi_n\pbk}=\delta(\bk-\bq)\delta_{mn}$, 
where $\inn{f}{g}=\int_{\mathcal{D}} d\br\, \overline{f}(\br)g(\br)$. Here,
$\delta(\bk)$ is the Dirac delta distribution, $\delta_{mn}$ is the Kronecker
delta, and $\overline{f}$ is the complex conjugate of $f$. The domain of
integration $\mathcal{D}$ is the periodic unit for the functions being
integrated; for Bloch orbitals, $\mathcal{D} = \mathbb{R}^3$. The $\ket{u_n\pbk}$
are orthonormal in the band index $n$ at a fixed $\bk$-point in reciprocal space:
that is, $\inn{u_m\pbk}{u_n\pbk}=\delta_{mn}$, where the inner product integrates
over one unit cell. Note that we assume closed-shell systems in this work.

The single-particle density can be represented in real space by the occupied
Bloch orbitals as
\begin{equation}
    \rho_s(\br) = \sum_n^{\text{occ}} \frac{V}{(2\pi^3)}
        \int_{\text{BZ}} d\bk\, \abs{\psi_n\pbk(\br)}^2,
\end{equation}
where $V$ is the volume of the unit cell and the integral is over the first
Brillouin zone. Since the Hamiltonian is diagonal in $\bk$, we can solve for the
Bloch orbitals in reciprocal space, requiring diagonalization only in one unit
cell. In practice, the Brillouin zone is sampled with a finite number of points;
in this work we use a Monkhorst--Pack mesh centered at the origin $\Gamma$ of
the Brillouin zone. \cite{monkhorst_special_1976} Thus, integrals over the
Brillouin zone become equally weighted sums over the $\bk$-mesh
\begin{equation}
    \frac{V}{(2\pi)^3} \int d\bk\, f(\bk) \mapsto \frac{1}{N_k} \sum_{\bk} f(\bk),
\end{equation}
where $N_k$ is the number of $\bk$-points in the mesh. Using a Monkhorst--Pack
mesh centered at $\Gamma$ yields Bloch orbitals having the periodicity of an
unfolded supercell comprised of $N_k$ primitive unit cells; this supercell is
referred to as the Born--von Karman cell. \cite{ashcroft_solid_2003} The Bloch
orbitals then obey the normalization convention
$\inn{\psi_m^{\bq}}{\psi_n\pbk}=N_k \delta_{\bk \bq}\delta_{mn}$, where the
integral is over the Born--von Karman cell.

\section{Methods}
The LOSC method consists of two steps. First, we find orbitals that are
spatially localized while remaining associated with specific energy ranges.
Next, we compute a curvature matrix modeling the magnitude of the deviation from
linearity. This is combined with the fractionally occupied localized orbitals to
correct the convex deviation of $E(N)$ from linearity at non-integer $N$, as
well as incorrect total energies at integer $N$. Both steps are implemented as
post-processing after a converged self-consistent field calculation.

\subsection{Localization}
The wave-like nature of the Bloch orbitals prohibits them from being spatially
localized. In order to obtain a state that is localized in space, the discrete
Fourier transform of the Bloch orbitals is used to produce Wannier functions
\cite{wannier_structure_1937}
\begin{equation}
\begin{split}
\ket{w_n^{\bR}} &= \frac{1}{N_k} \sum_{\bk} e^{-i\bk\cdot\bR} \ket{\psi_n^{\bk}}.
\end{split}
\end{equation}
Wannier functions inherit the periodicity of the Born--von Karman cell, and are
indexed by electron bands $n$ and unit cells $\bR$ in the supercell. They are
symmetric under translation by unit cell vectors, so that
$w_n^{\bR}(\br)=w_n^{\bZ}(\br-\bR)$; $\bR=\bZ$ is referred to as the home unit
cell. There is a unitary, or gauge, freedom at each $\bk$-point in the choice of
Bloch orbitals that comprise a Wannier function, so we can define
\emph{generalized} Wannier functions \cite{cloizeaux_orthogonal_1963} 
\begin{equation}
\begin{split}
\ket{w_i^{\bR}} 
    &= \frac{1}{N_k} \sum_{\bk} e^{-i\bk\cdot\bR} 
       \sum_n U_{ni}^{\bk} \ket{\psi_n^{\bk}} \\
    &= \frac{1}{N_k} \sum_{\bk} e^{-i\bk\cdot\bR} \ket{\phi_i\pbk},
\end{split}
\end{equation}
where we refer to $\ket{\phi_i\pbk}$ as a transformed Bloch orbital (TBO).
From now on, we will refer to generalized Wannier functions as Wannier
functions.

The gauge freedom $U^{\pbk}$ in the TBOs can be chosen such that the resulting
set of Wannier functions have advantageous properties. In order to obtain
localization in space, Marzari and Vanderbilt suggested choosing $U^{\pbk}$ that
minimize the Wannier functions' spatial variance
$\evo{\Delta r^2}_i = \evo{r^2}_i - \evo{\br}^2_i$, where
$\evo{x}_i = \evt{x}{w_i^{\bZ}}$; the resulting orbitals are called maximally
localized Wannier functions. \cite{marzari_maximally_1997} In molecules, the
scheme of minimizing spatial variance is referred to as Foster--Boys
localization. \cite{foster_canonical_1960} However, constructing maximally
localized Wannier functions from both valence and conduction bands is 
physically ill-motivated. Because bands far apart in energy can mix freely and
the Bloch bands form a complete basis, adding more virtual bands will result in
increasing spatial localization of the maximally localized Wannier functions,
with a corresponding loss of information about the energy dispersion of the
bands.

In order to preserve locality in energy while maintaining spatial localization,
enabling simultaneous treatment of the occupied and unoccupied spaces, we choose
the Wannier gauge that minimizes a cost function considering both energy and
spatial variance:
\begin{equation} \label{eqn:L2Cost}
F = (1-\gamma)\sum_i \evo{\Delta r^2}_i
+\gamma C \sum_i \evo{\Delta h_\text{s}^2}_i,
\end{equation}
where $0 \leq \gamma \leq 1$ and in the units used here
$C = \SI{1}{\bohr^2/\electronvolt^2}$, where $\si{\bohr}$ is the Bohr radius.
This cost function was first proposed by
\citet{gygi_computation_2003} for computations sampling the Brillouin zone only
at $\Gamma$ and implemented for such systems by \citet{giustino_mixed_2006}.
It was used for LOSC in molecules \cite{su_preserving_2020} to treat system
symmetries and degeneracies more robustly than the original localization, which
used soft energy windows; \cite{li_localized_2018} in molecular LOSC, the
localized orbitals are called \emph{orbitalets}. We recently extended $F$ to
systems with $N_k \geq 1$; \cite{mahler_wannier_2022} we refer to such orbitals
as dually localized Wannier functions (DLWFs). These formulations show how the
combination of occupied and unoccupied spaces can be localized simultaneously 
to produce Wannier functions that are localized in both space and energy. This 
construction is critical for addressing delocalization error in finite systems
because it allows for dynamic localization in the resulting orbitals; the
orbitals can qualitatively and quantitatively differ depending on the geometry
of the system. \cite{li_localized_2018, su_preserving_2020} In keeping with the
principle of universality in functional development, we use the same mixing
parameter in Eq.\ (\ref{eqn:L2Cost}), setting $\gamma=0.47714$. The value of
$\gamma$ has important implications for the LOSC method; see Section VI of the
Supplementary Information of \citet{su_preserving_2020}. Setting $\gamma = 0$,
for instance, yields maximally localized Wannier functions,
\cite{marzari_maximally_1997} while $\gamma = 1$ yields DLWFs that are pure
Fourier transforms (up to $\bk$-dependent phases) of the Kohn--Sham bands.

Note that we have not proven that a unique global minimum of $F$ exists; in
practice, $F$ has a fairly rugged landscape of solutions, and we have observed
multiple local minima. We choose the DLWFs yielding the smallest total cost.
Different DLWFs can produce somewhat different sLOSC corrections, with eigenvalues
varying by up to a few tenths of an \si{\electronvolt}. The problem of multiple
minima of $F$ has also been observed in molecular LOSC, \cite{mei_libsc_2022}
but was not found to be the dominant source of error. An additional question
worth exploring is the effect of symmetry breaking, such as that due to
perturbations of the crystal lattice, on the localization procedure.

The compromise between spatial and energy localization and the inclusion of
unoccupied orbitals are key to producing localized orbitals that can address
delocalization error while retaining size-consistency. For example, the DFA HOMO
and LUMO of H$_2^+$ at (or near) the dissociation limit are delocalized over the
whole molecule; since they are (nearly) degenerate, there exists a unitary
freedom in the subspace spanned by both.  Due to the symmetry of the system, we
expect to obtain two separate H$^{0.5+}$ fragments; the (small or) vanishing gap
means that any choice of $\gamma<1$ in Eq.\ (\ref{eqn:L2Cost}) will result
in half-occupied orbitals localized on each H atom. This is the physical
motivation for a localization scheme that minimizes the spatial variance of
occupied and unoccupied orbitals while allowing only orbitals that are close
in energy to mix. \cite{li_localized_2018}

\subsection{Energy corrections}
The deviation from energy linearity with respect to fractional charges is
characteristically quadratic in most exchange-correlation functionals.
\cite{zheng_improving_2011,hait_delocalization_2018,li_localized_2018} To
restore compliance with the PPLB condition for small finite systems, the
global scaling correction (GSC) was developed. GSC corrects the total energy
by an amount quadratic in the occupation numbers of the canonical molecular
orbitals. \cite{zheng_improving_2011,mei_exact_2021} This method is
effective at correcting the systematic deviation from the PPLB condition for
systems with fractional charges and leads to accurate prediction of
quasiparticle energies as the eigenvalues from the resulting one-electron
Hamiltonian. However, GSC is applicable only for systems of small and
moderate size; the convex deviation of conventional DFAs from the piecewise
linearity prescribed by the PPLB condition decreases with increasing system
size, and the delocalization error manifests instead as underestimated
ground-state energies for integer systems and incorrect linear
$E_{\text{gs}}(N)$ curves with wrong slopes at the bulk limit.
\cite{mori-sanchez_localization_2008} The localized orbital scaling
correction (LOSC) applies its energy correction adaptively by the
construction of localized orbitals, allowing systematic and size-consistent
correction of delocalization error.
\cite{li_localized_2018,su_preserving_2020} In this section, we discuss the
extension of LOSC to periodic systems.

A basic quantity in LOSC is the density matrix in the basis of DLWFs; its
elements are occupations
\begin{equation}
    \lambda_{ij}^{\bT\bR} = \mel{ w_i^{\bT} }{\rho_s}{ w_j^{\bR} }.
\end{equation}
The occupations between all pairs of DLWFs are used to remove quadratic
deviations, while the diagonal terms are used to restore linearity. The energy
correction defined by LOSC for each unit cell is given by
\begin{equation} \label{eqn:dELoscPBC}
    \Delta E^{\text{LOSC}} = \frac{1}{2N_k} \sum_{\bT \bR} \sum_{ij} 
    \widetilde{\kappa}_{ij}^{\bT\bR} {\lambda}_{ij}^{\bT\bR}
        ( \delta_{ij}^{\bT\bR} - \overline{\lambda}_{ij}^{\bT\bR} ),
\end{equation}
where $\delta_{ij}^{\bZ\bR}=\delta_{ij}\delta_{\bZ\bR}$; $\widetilde{\kappa}$
models the curvature of the deviation from linearity.

The diagonal terms in the energy correction are proportional to
$\lambda_{ii}^{\bT\bR}-\abs{\lambda_{ii}^{\bT\bR}}^2$; thus, if a DLWF has
integer occupancy (implying $\lambda_{ij}^{\bT\bR} = 0$ whenever $i \neq j$ or
$\bT \neq \bR$), then the energy correction due to that DLWF will also be zero.

The matrix $[\lambda_{ij}^{\bT\bR}]$ of occupations between the DLWFs is the
discrete Fourier transform of the occupation matrix between the TBOs. As such,
it is positive semidefinite and Hermitian, and
\begin{equation}
    \tr_c [\lambda_{ij}^{\bT\bR}] = \frac{1}{N_k} \sum_{\bk} N_f\pbk,
\end{equation}
where $\tr_c$ denotes the trace per unit cell and $N_f\pbk$ is the number of
electrons below the Fermi energy at $\bk$.

Following \citet{su_preserving_2020}, the elements of the curvature matrix
are given by
\begin{equation} \label{eqn:kappa2}
\begin{split}
    \widetilde{\kappa}_{ij}^{\bT\bR} = & \erf(8 S_{ij}^{\bT\bR})
    \sqrt{ \kappa_{ii}^{\bT\bT} \kappa_{jj}^{\bR\bR}  } \\
    + &\erfc(8 S_{ij}^{\bT\bR})
    {\kappa}_{ij}^{\bT\bR}.
\end{split}
\end{equation}
Here, $\erfc(r) = 1 - \erf(r)$ is the complementary error function.
$S_{ij}^{\bT\bR}$ is the absolute overlap between DLWFs,
\begin{equation} \label{eqn:ovsmetric}
    S_{ij}^{\bT\bR} = \int d\br\, 
    \sqrt{ \rho_i^{\bT}(\br) \rho_j^{\bR}(\br) },
\end{equation}
where $\rho_i^{\bT}(\br) = \abs{w_i^{\bT}(\br)}^2$ is a DLWF's charge density.
The matrix elements ${\kappa}_{ij}^{\bT\bR}$ in Eq.\
(\ref{eqn:kappa2}) are given by 
\begin{equation} \label{eqn:kappa1}
    \kappa_{ij}^{\bT\bR} = J[\rho_i^{\bT},\rho_j^{\bR}] - 
        X[\rho_i^{\bT},\rho_j^{\bR}],
\end{equation}
with
\begin{subequations}
\begin{align}
    J[\rho_i^{\bT},\rho_j^{\bR}] &= \iint d\br\, d\br'\,
    \rho_i^{\bT}(\br) \rho_j^{\bR}(\br') K(\abs{\br-\br'}),
        \label{eqn:kappaJ} \\
    X[\rho_i^{\bT},\rho_j^{\bR}] &= \tau \frac{2C_\text{X}}{3} \int d\br\,
    \Big[ \rho_i^{\bT}(\br)\rho_j^{\bR}(\br) \Big]^{2/3}. \label{eqn:kappaX}
\end{align}
\end{subequations}
In the above, $K(r)=1/r$ is the Coulomb kernel,
$C_{X}=\frac{3}{4}(\frac{6}{\pi})^{1/3}$ is the Dirac exchange constant,
\cite{dirac_note_1930} and $\tau=6(1-2^{\text{-}1/3})\approx 1.2378$ is a
nonempirical parameter.\cite{li_localized_2018} The derivation of how this
correction restores the PPLB condition can be found in the supplementary data of
\citet{li_localized_2018}. The use of $\widetilde{\kappa}$ instead of $\kappa$
was introduced because the cost function in Eq.\ (\ref{eqn:L2Cost}) can induce
discontinuous jumps between localization characters during molecular
dissociation.\cite{su_preserving_2020} The diagonal elements of
$\widetilde{\kappa}$ and $\kappa$ are equal, so when $\lambda_{ii} \in \{0,1\}$
the corrections from $\widetilde{\kappa}$ and $\kappa$ are the same. In
practice, $X[\rho_i^{\bT},\rho_j^{\bR}]$ term is evaluated using numerical
integration on a grid of real-space points. The Coulomb term
$J[\rho_i^{\bT},\rho_j^{\bR}]$ is evaluated in a plane wave basis, detailed in
Sec.\ \ref{sec:coulomb}.

Applying the extension of Janak's theorem\cite{janak_proof_1978} to the
generalized Kohn-Sham theory, \cite{cohen_fractional_2008} the LOSC energy
correction of Eq.\ (\ref{eqn:dELoscPBC}) yields corrections to the Bloch orbital
energy eigenvalues $\epsilon_n\pbk$ given by
\begin{multline} \label{eqn:depsLosc}
\Delta\epsilon_{n}^{\bk} = \sum_{i}
\widetilde{\kappa}_{ii}^{\bZ\bZ} 
\left(\frac{1}{2} - \lambda_{ii}^{\bZ\bZ} \right)
\abs{U_{ni}^{\bk}}^2 \\
- \sum_{\bR i\neq \bZ j} 
\widetilde{\kappa}_{ij}^{\bZ\bR} \Re \left\{
\lambda_{ij}^{\bZ\bR} 
e^{i\bk \cdot\bR }\overline{U}_{ni}^{\bk}U_{nj}^{\bk} \right\}.
\end{multline}
Consider the diagonal corrections given by the first summand in Eq.\
(\ref{eqn:depsLosc}). There is no correction to the eigenvalue when a DLWF is
half-occupied ($\lambda_{ii} = \frac12$); on the other hand, the correction is
maximal when it is completely occupied or unoccupied. \citet{li_localized_2018}
observed that the slopes of the quadratic DFA and the correct linear $E(N)$
curves agree at half-integer $N$. 
Since the frontier orbital energy corresponds to this slope, we see that
accurate frontier orbital energies are given by half-occupied frontier orbitals.
The LOSC correction to the orbital energies arrives naturally at this
conclusion, additionally agreeing with Slater transition state theory.
\cite{slater_quantum_1974,slater_quantum_1974-1}

We may also view LOSC as a correction to the Kohn--Sham Hamiltonian. It is given
by the functional derivative of the energy correction with respect to the density
operator under the frozen orbital approximation,
\begin{equation}
    \Delta v = \left. \frac{\delta \Delta E^{\text{LOSC}}}{\delta \rho_s}
           \right\rvert_{\{w_i^{\bR}\}}, \\
\end{equation}
and can be written in operator form as
\begin{equation}
    \Delta v = \frac12 \sum_{ij, \bT \bR} \widetilde{\kappa}_{ij}^{\bT\bR}\,
    \left(\frac{\delta_{ij}^{\bT\bR}}{2} - \lambda_{ij}^{\bT\bR} \right) 
    \outt{ w_i^{\bT} }{ w_j^{\bR} } + \text{h.c.}
\end{equation}
(See the Supplemental Material for details on this derivation.) The correction
to the $n^{\text{th}}$ Bloch orbital eigenvalue is then given by
$\Delta\epsilon_{n}^{\bk} = \mel{\psi_n\pbk}{\Delta v}{\psi_n\pbk}$.

In practice, the energy corrections are applied to disentangled Bloch orbitals.
The conduction bands of most systems cannot be formed into sets of bands that do
not cross anywhere in the Brillouin zone, a condition referred to as band
entanglement. In order to obtain a finite set of bands for localization and
energy correction, we use the disentanglement procedure outlined by
\citet{souza_maximally_2001} This procedure obtains $N_w$ bands from a
set of $N_b \geq N_w$ Bloch orbitals at each $\bk$-point, chosen such that the
subspace spanned by the disentangled bands is as smooth as possible in $\bk$.
To correct the band gap of semiconductors and insulators, we include sufficiently
many virtual bands in the construction of the Wannier functions to converge the
localization of the frontier bands (that is, the valence band maximum and
conduction band minimum). \cite{mahler_wannier_2022} We find that
$N_b = N_{\text{occ}} + 3 N_{\text{coord}}$ and 
$N_w = N_{\text{occ}} + 2 N_{\text{coord}}$, where $N_{\text{occ}}$ is the
number of occupied bands per unit cell and $N_{\text{coord}}$ is the
coordination number of the lattice, are sufficient; see the Supplemental
Material for details. The $N_w$ disentangled Bloch bands yields $N_w$ DLWFs per
unit cell, so there are $N_w N_k$ DLWFs in the Born--von Karman supercell on
which they are periodic. The energy corrections for the $N_w$ disentangled Bloch
orbitals at each $\bk$-point are implemented using Eq.\ (\ref{eqn:depsLosc}).

In this work, we restrict our attentions to closed-shell systems. Extending
the LOSC method to spin-polarized materials is accomplished by finding the
corrections from the spin-up and spin-down DLWFs independently and
summing them to obtain $\Delta E^{\text{LOSC}}$; this functionality is planned
for the next version of LOSC. However, treating the strong correlation common to
open-shell materials brings its own set of challenges beyond the scope of this
work. We discuss them briefly in Sec.\ \ref{sec:discussion} below.

\subsection{Coulomb integrals} \label{sec:coulomb}
Accurate calculation of  the Coulomb interaction $J[\rho_i^{\bT},\rho_j^{\bR}]$
is needed for LOSC to restore the PPLB condition.  In PBCs, a plane-wave basis
is typically employed. The Coulomb energy is diagonal in this basis, and the
double integral required in real space collapses to a single sum over basis
vectors $\bG$:
\begin{equation}
    J[\rho_i^{\bT},\rho_j^{\bR}] = \sum_{\bG} \frac{4\pi}{G^2}\, \conj{\rho}_i^{\bT}(\bG) \rho_j^{\bR}(\bG),
\end{equation}
where $G = \abs{\bG}$. However, this sum converges only for neutral charge
distributions, for which the $\bG = \bZ$ term vanishes. The DLWF densities are
individually charged, so ignoring the divergent term coming from the net charge
will significantly underestimate the Coulomb energy. There are many methods to
evaluate the Coulomb energy for charged densities in the plane-wave basis
accurately, including those of
\citet{makov_periodic_1995,kantorovich_elimination_1999,dabo_electrostatics_2008},
and \citet{li_electronic_2011}.
We choose the spherical cutoff method,
\cite{onida_ab_1995,jarvis_supercell_1997,rozzi_exact_2006} truncating the
Coulomb kernel in Eq.\ (\ref{eqn:kappaJ}) at a cutoff radius $R_c$; this is taken to
be half the length of the shortest Born--von Karman supercell lattice vector,
ensuring that the Coulomb interactions between the a DLWF density and its images
in neighboring supercells are zero. Thus, the spherical cutoff Coulomb kernel is
\begin{equation} \label{eqn:VSphCut}
K_c(r; R_c)=
\begin{cases} 
    1/r & r <  R_{c} \\
    0 & r \geq R_c,
\end{cases}
\end{equation}
which has Fourier coefficients
\begin{equation} \label{eqn:VGSphCut}
K_c(G; R_c)=
\begin{cases} 
    \dfrac{4\pi}{G^2} [1-\text{cos}(G R_c)] & G \neq 0 \\[0.6em]
    2\pi R_c^2 & G = 0.
\end{cases}
\end{equation}
Observe that $K_c(G;R_c)$ does not diverge for any $\bG$. As long as the pair
of DLWF densities in Eq.\ (\ref{eqn:kappaJ}) lie in a sphere of radius $R_c$,
the spherical cutoff method is also accurate in highly anisotropic unit cells,
unlike schemes such as that of Makov and Payne.
\cite{kantorovich_elimination_1999} We enforce this containment condition in
practice by checking that each DLWF density is well contained in a volume
spanned by half of each Born--von Karman supercell lattice vector, and only
compute curvature elements between pairs of DLWF densities that have centers
closer together than $R_c$. We evaluate the Coulomb integrals on the unfolded
supercell in the plane-wave basis, which requires a fast Fourier transform (FFT)
of the DLWF densities on the supercell. 

\subsection{Screening}
Applying LOSC with a bare Coulomb interaction leads to severe overcorrection of
semiconductors' band gaps in PBCs. However, this is not surprising; we
anticipate an effect of the other electrons in the lattice on
$J[\rho_i^{\bT},\rho_j^{\bR}]$. As shown by highly accurate methods such as
$GW$, a screened Coulomb interaction is required to model the interaction
between electrons in a periodic system accurately.
\cite{hedin_new_1965,martin_interacting_2016} Recently, Mei and coworkers also
found that the deviation from linearity of the total energy as a function of
canonical orbital occupations is given to second order by a screened
interaction. \cite{mei_exact_2021} We model the screening phenomenologically,
attenuating the long-range $1/r$ behavior of the spherical cutoff Coulomb
interaction by a complementary error function. This modifies the Coulomb kernel
to read
\begin{equation} \label{eqn:VScreenReal}
K_s(r; R_c, \alpha)=
\begin{cases} 
    \erfc (\alpha r)/r & r < R_{c} \\
    0& r \geq R_c,
\end{cases}
\end{equation}
where $\alpha$ is a screening parameter. We choose the $\alpha$ that best
reproduces the experimental band gaps of a test set of semiconductors and
insulators. For $r$ larger than the screening radius $\alpha^{-1}$,
$K_s(r; R_c, \alpha)$ decays exponentially instead of as $1/r$.  The Fourier
coefficients of $K_s$ are
\begin{widetext}
\begin{equation} \label{eqn:VScreen}
K_s(G; R_c, \alpha)=
\begin{cases} 
\dfrac{4\pi}{G^2}\left[ 1 - \cos(G R_c) \erfc(\alpha R_c)
-e^{-(G/2\alpha)^2} \Re \left\{ \erf\left( 
\alpha R_c +\dfrac{i G}{2\alpha} \right) \right\} \right] & G \neq 0 \\[1.1em]
2\pi R_c^2 + \pi \erf (\alpha R_c) \left(\alpha^{-2} - 2 R_c^2\right) - 2\sqrt{\pi} e^{-(\alpha R_c)^2}R_c /\alpha &  G = 0.
\end{cases}
\end{equation}
\end{widetext}
The error function is unbounded for complex arguments, overflowing
double-precision floating-point numbers even for relatively small $G$. Thus,
we evaluate $K_s(G; R_c, \alpha)$ with a scaled form of $\erf z$ called the
Faddeeva function, implemented in the numerically stable ACM Algorithm 916.
\cite{zaghloul_algorithm_2012, zaghloul_remark_2016} For details, see the
Supplemental Material.

In principle, the screening is system-dependent. Improved accuracy would be
attainable by setting its value to best reproduce each material's band gap.
However, the phenomenological screening model of sLOSC does not enable doing so
while retaining predictive ability. This would require a rigorously screened
Coulomb (or Hartree-exchange-correlation) interaction based on the linear
response function $\chi(\br,\br') = \delta \rho_s(\br)/\delta v(\br')$.
\textit{Ab initio} screening of this kind appears in the extensions of the GSC
method to hybrid functionals \cite{zheng_a_2013} and in the following
exploration of orbital relaxation on GSC, \cite{zhang_orbital_2015}, the GSC2
method, \cite{mei_exact_2021} as well as in recent work on Koopmans-compliant
functionals.
\cite{colonna_screening_2018, nguyen_koopmans_2018, colonna_koopmans_2022}
For small, finite systems, the delocalization error is quantified by    
$\partial^2 E/\partial n_i^2$, where $n_i$ is the occupation number of the
Kohn--Sham orbital $\ket{\psi_i}$; analytical expressions for
$\partial^2 E/\partial n_i^2$ were derived in \citet{yang_analytical_2012}
In sLOSC, linear-response screening would very likely increase the accuracy,
but at substantial computational cost to compute the nonlocal $\chi(\br,\br')$.

\section{Results} \label{sec:results}
We use the PBE functional \cite{perdew_generalized_1996} for the parent
DFA calculations, with optimized norm-conserving Vanderbilt pseudopotentials
\cite{hamann_optimized_2013} generated by PseudoDojo.
\cite{van_setten_pseudodojo_2018} Both self-consistent field (SCF) and non-SCF
calculations are carried out in the \texttt{Quantum ESPRESSO} code suite.
\cite{giannozzi_quantum_2009,giannozzi_advanced_2017}

The energy cutoff for the fast Fourier transform is set to \SI{100}{\rydberg}
for wavefunctions and \SI{400}{\rydberg} for densities. The Brillouin zone is
sampled with Monkhorst--Pack meshes centered at $\Gamma$, which is necessary
for the Wannier functions to be periodic on the Born--von Karman supercell. For
SCF calculations, we use a \kmt{16} $\bk$-mesh, while the other calculations
are performed on \kmt{6} grids. The localization step of LOSC is implemented in
a modified fork of the \texttt{wannier90} code,
\cite{mostofi_updated_2014,pizzi_wannier90_2020} and the energy correction as
module to a fork of \texttt{Quantum ESPRESSO}.

\begin{figure}[!htp]
\centering
\includegraphics[scale=0.065]{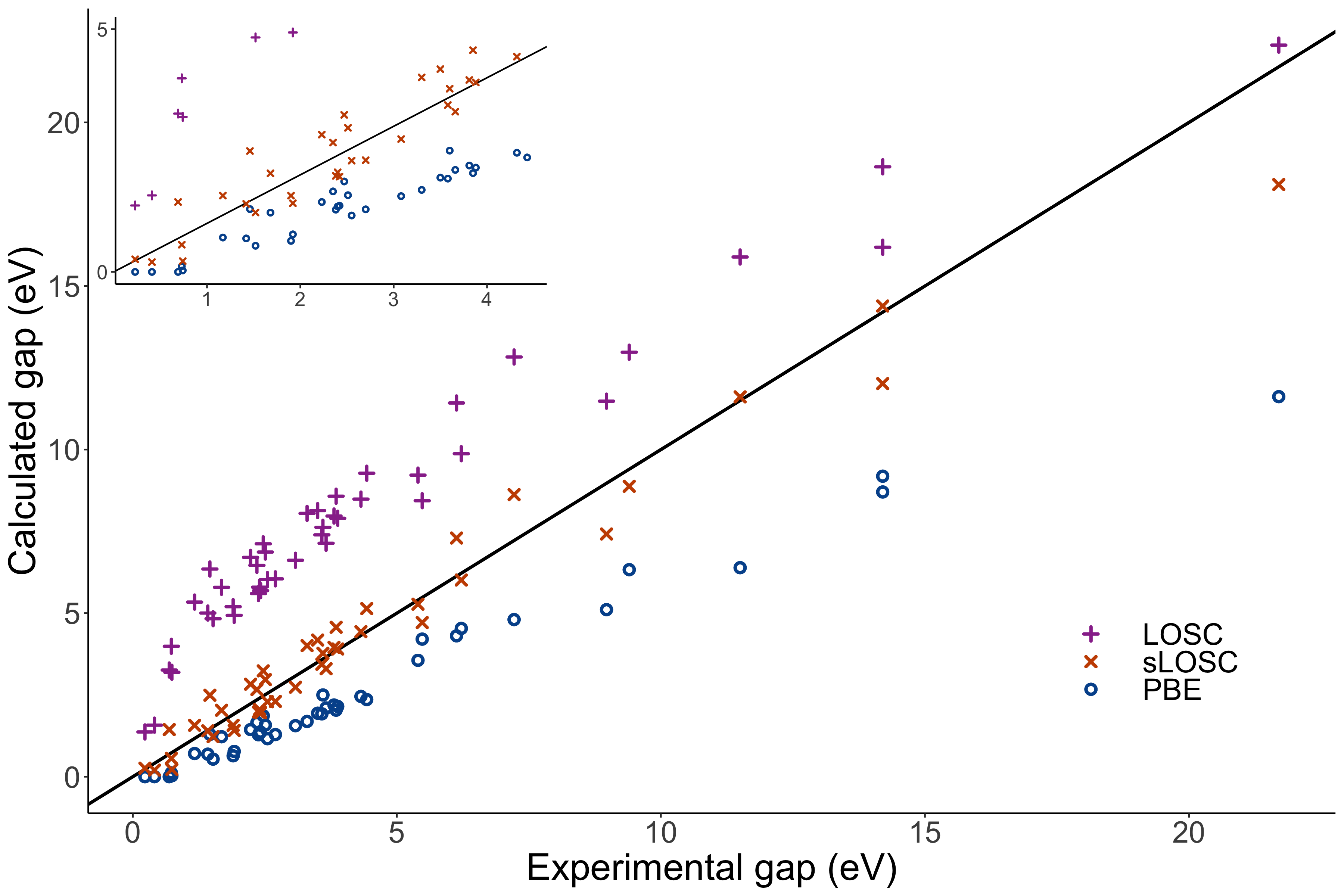}
\caption{Comparison of experimental band gaps with those calculated by PBE
         (\texttt{o}), sLOSC ($\boldsymbol{\times}$), and unscreened LOSC
         (\texttt{+}). The inset shows systems with an experimental band gap
         less than \SI{5}{\electronvolt}. } 
\label{fig:scatter_gaps}
\end{figure}

To determine an optimal screening parameter $\alpha$, we minimize the mean
absolute percent error (MAPE) on the SC/40 set of semiconductors with
experimentally available band gaps \cite{heyd_energy_2005} together with six
additional large-gap insulators. The experimental band gaps studied range from
\SIrange{0.23}{21.7}{\electronvolt}. We find that $\alpha =$
\SI{0.15}{\bohr^{-1}} achieves the lowest MAPE; coincidentally, this value is
numerically equal to the screening parameter used in the HSE density functional.
\cite{heyd_energy_2005} As shown in Fig.\ \ref{fig:scatter_gaps}, LOSC with Coulomb
screening (sLOSC) yields marked improvement of the band gap for the test set in
comparison with the parent functional. It is also apparent that unscreened LOSC
overcorrects the band gaps; indeed, it is less accurate than the parent functional.
The performance of sLOSC in molecules is better than the parent functional, but
unscreened LOSC achieves the best performance in molecular systems
(see Table \ref{tab:mae}). The Supplemental Material details the variation in
performance of screened LOSC for both bulk systems and molecules with the 
creening parameter $\alpha$.

\begin{table}[H]
\caption{Mean absolute percent error of the band gap for PBC and molecular test
         sets. For details on the systems tested, see the Supplemental Material.
         }
\label{tab:mae} 
    \centering
    \begin{ruledtabular}
    \begin{tabular}{@{}r|rrr@{}}
        Method & PBE & LOSC & sLOSC \\[0.25em]
        \hline \\[-0.75em]
        PBC & $47.5\%$ & $158.6\%$ & $19.7\%$ \\
        Molecule & $79.8\%$ & $10.1\%$ & $43.6\%$
    \end{tabular}
    \end{ruledtabular}
\end{table}

\begin{figure}[!htp]
\centering
\includegraphics[scale=0.55]{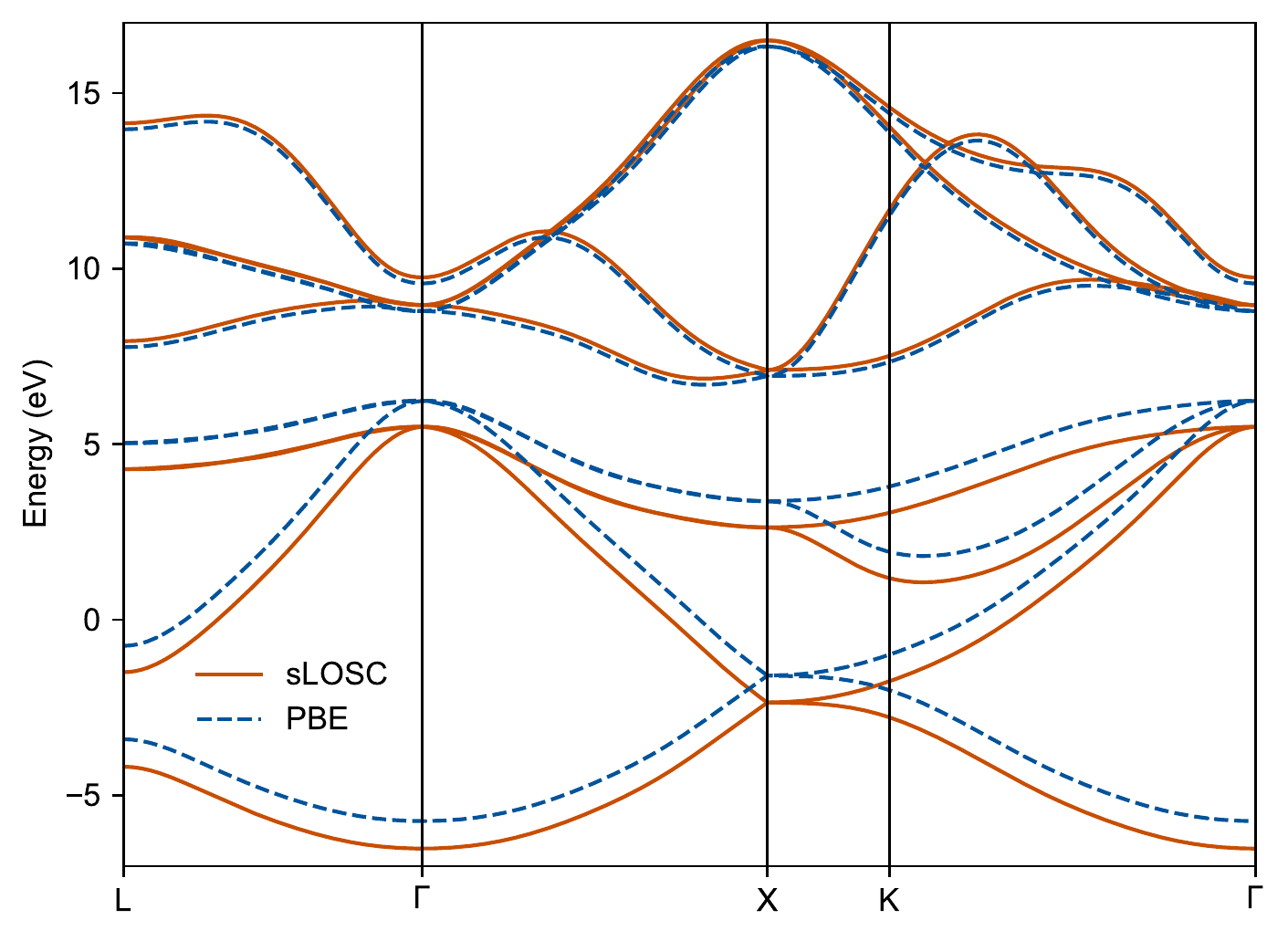}
\caption{Band structure of silicon under the PBE functional (dashes) and sLOSC
         (solid). The Fermi energy of the PBE calculation was
         \SI{6.23}{\electronvolt}, while the PBE with LOSC Fermi energy was
         \SI{5.49}{\electronvolt}.} 
\label{fig:si_bands}
\end{figure}

\begin{figure}[!htp]
\centering
\includegraphics[scale=0.55]{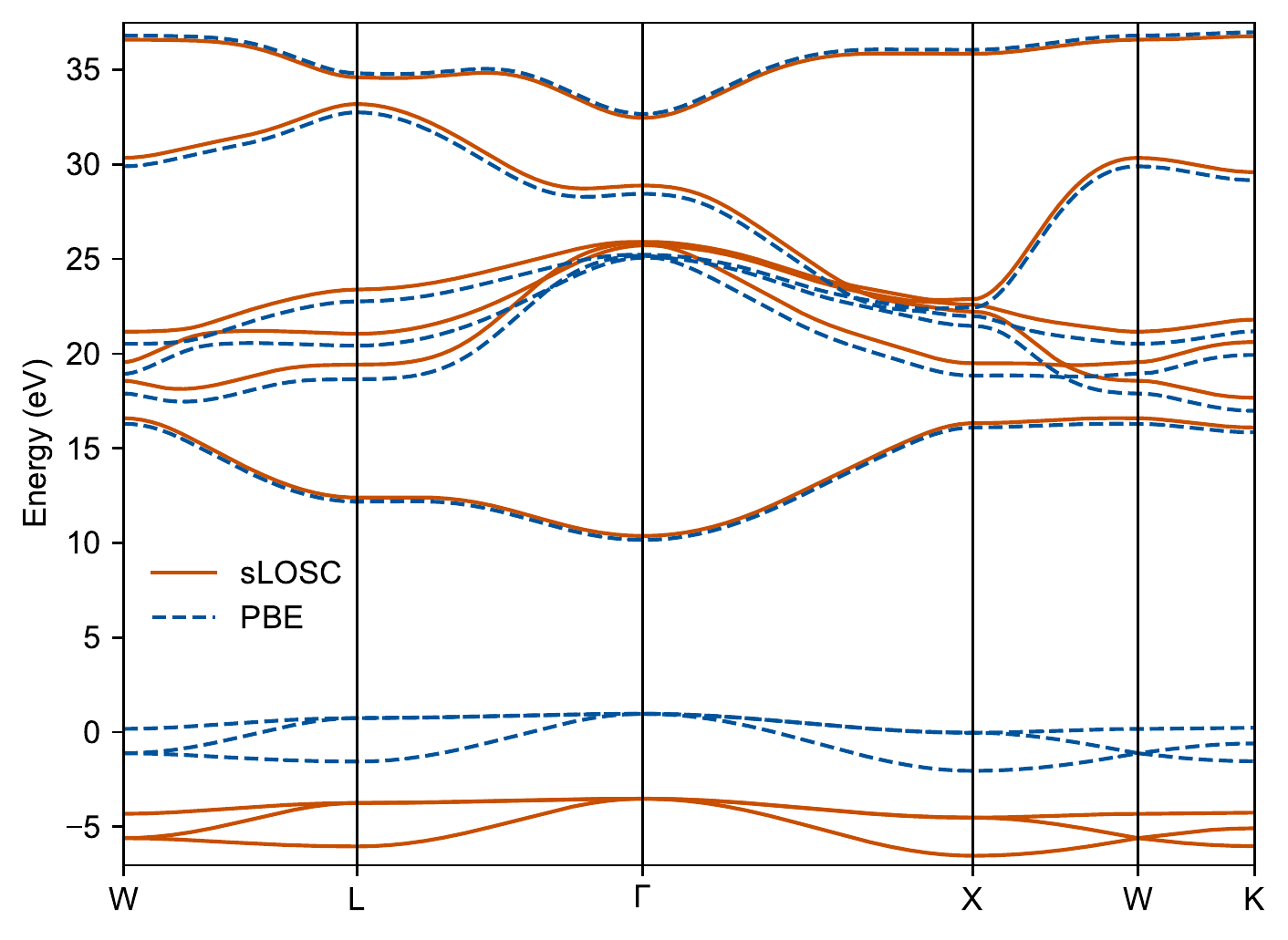}
\caption{Band structure of lithium fluoride under the PBE functional (dashes)
         and sLOSC (solid). The Fermi energy of the PBE calculation was
         \SI{0.97}{\electronvolt}, while the PBE with LOSC Fermi energy was
         \SI{-3.52}{\electronvolt}. The core states are not included in the
         figure.}
\label{fig:lif_bands}
\end{figure}

The band structures of sLOSC and of the parent functional are shown for the
small-gapped semiconductor silicon in Fig.\ \ref{fig:si_bands} and the
larger-gapped insulator lithium fluoride in Fig.\ \ref{fig:lif_bands}. They use
the disentangled band structures, which are numerically indistinguishable from
the true band structure at and below the conduction band minimum for the parent
functional. Wannier interpolation in the same DLWF basis as that used in sLOSC
is used to find the energy at the points in the Brillouin zone not explicitly
treated by the localization and energy correction. The sLOSC correction to the
band structure comes largely from the more localized DLWFs, for which there
is a larger Coulomb self-energy $J[\rho_i^\bR,\rho_i^\bR]$. Because of this,
sLOSC mostly corrects the energy of the occupied bands (which we observe to
correspond closely to the occupied DLWFs in semiconductors); it affects the
virtual bands much less. More work on molecule-surface and surface-surface
interactions is required to determine whether LOSC yields correct energy level
alignment and whether the larger correction to the valence bands is physically
meaningful.

\section{Discussion} \label{sec:discussion}
We have shown that despite the simple form of its phenomenological Coulomb
screening, sLOSC systematically corrects the band gap error associated with the
parent functional for materials spanning a large range of band gaps. Screening
improves the correction of delocalization error in bulk systems, but degrades
the accuracy of molecular systems' band gaps relative to unscreened LOSC;
however, sLOSC still offers better band gaps than those computed by the parent
functional. One key remaining challenge is to model the curvature more
accurately for all systems; we expect that linear response of the electron
density, used by \citet{mei_exact_2021} for accurate screening of the 
Kohn--Sham orbitals, could be used to find the exact expression for
$\partial^2 E / \partial \lambda_{ij}$. This would alleviate the error imposed
by modeling $\kappa$ as a difference between Coulomb repulsion and Dirac 
exchange.

We implement the energy correction as a post-processing step to a
self-consistent calculation; such corrections are accurate when the change in
electron density is small and hence the total energy correction is small. For
every system considered in this work, $\Delta E^{\text{LOSC}}$ does not exceed
three parts in $10^5$. The corresponding change to the density for such systems
is also expected to be minimal. LOSC can also be implemented
self-consistently; \cite{mei_self-consistent_2020} this can correct the
delocalization error of the total density, improving the accuracy of LOSC for
systems with large total energy corrections. A self-consistent implementation
of sLOSC could be necessary for the accurate computation of heterogeneous and
interfacial systems. Since delocalization error leads to incorrect charge
distributions, the sLOSC energy correction is likely to be larger, and
self-consistently correcting the delocalized density is expected to yield
better orbital energies.

Work is ongoing to implement sLOSC for spin-polarized materials and to
investigate its treatment of metals. The DLWFs of gapless systems constructed
from Bloch bands near the Fermi energy are expected to have occupations
$\lambda_{ii}$ close to $\frac12$, which means that the sLOSC correction to
those eigenvalues will be small. While there may be changes to the overall
band structure, it is likely that such systems will remain gapless.

This may not hold in semimetals, whose valence and conduction bands
cross only in a small volume (or a single point) of the Brillouin zone.
Metals, on the other hand, have one band that crosses the Fermi energy.
sLOSC can open a gap in systems the DFA predicts to be semimetals; this
occurs with the smallest-gapped system in our test set, InSb. Thus, it is
not certain that true semimetals would remain so after the sLOSC correction.
In addition, the treatment of strong correlation due to (near) degeneracy of
spin states and the inclusion of topological or spin-orbit effects are
beyond the scope of the current work. A modification of molecular LOSC to
include fractional spins was developed in \citet{su_describing_2018}; it
could possibly be extended to bulk materials as well.

\subsection{Comparison with other methods}
\subsubsection{DFT+U(+V)}
(s)LOSC is related to the DFT+U
\cite{anisimov_band_1991, cococcioni_linear_2005} method for correcting
delocalization error. The kinship can be seen in the similarity of the sLOSC
energy correction, Eq.\ (\ref{eqn:dELoscPBC}), to the rotationally invariant
DFT+U correction 
\cite{oregan_subspace_2012}
\begin{equation}
    \Delta E_{\text{DFT+U}} = \frac12 \sum_{\ell,\sigma} 
        \tr \left[\boldsymbol{n}_{\ell \sigma}
                  (1 - \boldsymbol{n}_{\ell \sigma})\right] U_\ell,
\end{equation}
where $U_\ell$ is the effective Hubbard parameter for orthonormal local orbital
(LO) $\ell$, combining atom and orbital indices, and
$\boldsymbol{n}_{\ell\sigma}$ is the LO occupation matrix. Both offer an
adjustment to the total energy quadratic in the occupation of the LOs.
The energy correction of DFT+U comes only from interactions between LOs on
the same atom, although DFT+U+V \cite{campo_extended_2010} extends this to
interactions between atoms, analogous to the off-diagonal curvature elements
$\widetilde{\kappa}_{ij}$ of (s)LOSC. However, the LOs of DFT+U(+V) are static
(usually being $d$ and $f$ orbitals on transition metal centers), while the
DLWFs of sLOSC dynamically localize based on the gauge set by the cost
function $F$. Thus, where DFT+U(+V) must recompute the effective Hubbard
parameter for every perturbation of the crystal structure or molecular
geometry, the size of the correction in (s)LOSC follows from the DLWFs.

It is worth noting that, while DFT+U(+V) does not explicitly include energy
localization in its construction, the Hubbard correction applies primarily to
the Kohn--Sham orbitals that have the most overlap with the LOs; viewed
another way, the (spatially) localized orbitals that have the most
energy-local character (via their large overlap with energy eigenstates).
\cite{himmetoglu_hubbard_2014} In particular, the $d$ and $f$ atomic orbitals
of transition metals correspond closely to flat bands in reciprocal space,
which carry some energy information implicitly. However, they are
independent of the of the system's geometry. In contrast, the LOs of LOSC are
dynamic:\ the orbitals can change with the geometric structure of the system.
This is key to their utility in finite systems. In compact structures near
equilibrium, the LOSC LOs can replicate the Kohn--Sham canonical orbitals, while
becoming localized as chemical bonds are stretched. This allows the LOSC total
energy correction to change with the geometry, as seen in
\citet{li_localized_2018} and \citet{su_preserving_2020}

Neither sLOSC nor DFT+U are suitable for solving the analogue of delocalization
error for systems with fractional spin.
\cite{cohen_fractional_2008, mori-sanchez_discontinuous_2009} For molecules,
fractional-spin LOSC (FSLOSC) \cite{su_describing_2018} extends the original LOSC
method to this case; the judiciously modified DFT (jmDFT) method
\cite{bajaj_communication_2017, bajaj_non-empirical_2019} does the same for
DFT+U.

\subsubsection{Koopmans-compliant functionals}
Koopmans-compliant functionals \cite{dabo_non-koopmans_2009, dabo_koopmans_2010}
mitigate delocalization error by enforcing the PPLB linearity condition directly:
in the Koopmans integral (KI) formulation, \cite{borghi_koopmans-compliant_2014} 
\begin{equation}
    \Delta E_{\text{KI}} = 
        \sum_i \alpha_i \left[ f_i \eta_i - 
            \int_0^{f_i} ds_i\, \evt{h_s(s_i)}{\phi_i} \right].
\end{equation}

Here $\alpha_i$ is an orbital-dependent screening function based on the
relaxation of the LOs $\phi_i$; $f_i$ is the (fractional) occupation of
$\phi_i$; $\eta_i = \int_0^1 ds_i\, \evt{h_{\text{PZ}}(s_i)}{\phi_i}$,
integrating the Perdew--Zunger self-interaction corrected Kohn--Sham
Hamiltonian, \cite{perdew_self-interaction_1981} gives the linearized slope of
the energy with respect to $f_i$; and the last term computes the nonlinearity in
$E$ that is replaced by $f_i \eta_i$. Like (s)LOSC, Koopmans-compliant
functionals are dependent on the choice of localized orbitals.
\cite{borghi_variational_2015} In extended systems, localized orbitals are
necessary for a Koopmans-compliant correction to have any effect,
\cite{nguyen_koopmans_2018} and screening has also been found to effect
improvements in band gap calculation.
\cite{colonna_screening_2018, colonna_koopmans_2022}

An advantage of sLOSC over the Koopmans-compliant functionals for extended
systems is that the DLWFs treat the valence and conduction bands together; on
the one hand, DLWFs are empirically robust to increasing the number of conduction
bands from which they are constructed, and on the other, sLOSC can in principle
be applied to metals without additional modification. For gapped systems, the
energy localization inherent in the DLWF cost function enforces separation between
the occupied and virtual electronic manifolds without manual input. The system
with the smallest gap in our analysis, indium antimonide (InSb, experimental gap
\SI{0.23}{\electronvolt}), which is predicted to be gapless by the DFA (and whose
sLOSC gap is \SI{0.260}{\electronvolt}) has DLWF occupations
$\lambda_{ii}^{\bR\bR} \geq 0.98$ in the valence manifold and
$\leq 0.0074$ in the conduction manifold. It is conceivable that DLWFs could also
serve as effective LOs for Koopmans-compliant methods, even if their subspaces
corresponding to the valence and conduction bands are not variational for total
Koopmans-compliant energy.

\subsubsection{Additional methods}
The Fermi--L\"owdin orbital (FLO) self-interaction correction (SIC), 
\cite{pederson_communication_2014, kao_self-consistent_2017,
      jackson_towards_2019} which has its roots in the Perdew--Zunger (PZ)
self-interaction correction method, \cite{perdew_self-interaction_1981} also
uses localized orbitals for an energy correction. However, the self-interaction
error treated by both FLOSIC and PZ-SIC is well-defined only for one-electron
systems; LOSC and its derivatives account for the many-electron nature of
delocalization error explicitly. \cite{mori-sanchez_many-electron_2006}

As mentioned in \citet{su_describing_2018}, the generalized transition state
method \cite{anisimov_transition_2005} and the Wannier-function method of Ma
and Wang \cite{ma_using_2016} are effective at improving band gap predictions.
However, since they do not mix valence and conduction bands to create
fractionally occupied orbitals, they cannot change the total energy of the DFA
calculation; thus, they cannot restore size-consistency to DFAs and will not be
able to capture (for instance) molecular dissociation at the same time as
improving band gap predictions. This problem is shared by early
Koopmans-compliant methods, which used the Kohn--Sham orbitals as the $\phi_i$;
it underlies the observation of \citet{nguyen_koopmans_2018} that localized
orbitals such as Wannier functions are required for Koopmans compliance in
extended systems.

\subsection{Computational efficiency} \label{sec:efficiency}
The sLOSC method as implemented in this work scales as $O(N_w^3 N_k)$ for the
localization step, $O(N_w^2 N_G)$ for the computation of curvature elements,
and $O(N_w N_G \log N_G)$ for the FFT of the DLWF densities. Here, $N_G$ is
the number of plane waves in the unfolded supercell, which is $N_k$ times the
number of plane waves in the unit cell. Calculating the curvature and energy
corrections is the computational bottleneck for the systems evaluated in this
work, with wall times for each system reaching a few hours using 16 threads on
an Intel Xeon E5-2630v3 processor. The systems that took the longest time were
those with the largest number of core states, which have no effect on frontier
state corrections; these could be neglected if only a correction to the band gap
is desired. The running time was divided fairly evenly between the computation
of the matrix elements defined in Eqs.\ (\ref{eqn:ovsmetric}, \ref{eqn:kappaJ},
\ref{eqn:kappaX}). We note that the size of the integration domain for these
quantities could be reduced from the full Born--von Karman supercell if the
relevant DLWF densities are contained in a smaller region. This is supported by
the fact that systems that had similar localizations for a \kmt{4} and \kmt{6}
$\bk$-mesh yielded very similar energy corrections.

Along the same lines, we find that for the systems in the test set the
correction results are converged with a \kmt{6} $\bk$-mesh, but this is only
necessary to achieve a converged localization. Certain systems exhibit a
qualitatively different localization with a smaller \kmt{4} $\bk$-mesh; however,
by decreasing the value of $\gamma$ in the localization cost function $F$, a set
of DLWFs qualitatively similar to the \kmt{6} case can be obtained.

Some other methods that attempt to address delocalization error in bulk
calculations, such as the approach of \citet{ma_using_2016} and the screened
range-separated hybrid functional, \cite{wing_band_2021} rely on supercell 
self-consistent calculations. These have cubic scaling in the number of 
electrons, so an unfolded Born--von Karman supercell arising from $N_k$
$\bk$-points sampling a unit cell with $N_w$ Wannier functions scales as
$O(N_k^3N_w^3)$. Both of the aforementioned methods use Wannier functions as a
localized charge representation and rely on manually choosing the Bloch orbitals
to comprise the Wannier functions representing the frontier of the occupied
space. The sLOSC method uses DLWFs, which naturally supply Wannier functions
representing the frontier of the occupied and unoccupied spaces without the need
for manual energy windowing. 

\begin{acknowledgments}
A.M., J.Z.W., N.Q.S., and W.Y. acknowledge support from the National Science
Foundation (Grant No.\ CHE-1900338); A.M. was additionally supported by the
Molecular Sciences Software Institute Phase-II Software
Fellowship, and J.Z.W. by the National Institutes of Health (Grant No.\
5R01GM061870).
\end{acknowledgments}

\bibliography{main}
\bibliographystyle{apsrev4-2}

\end{document}


\title{Supplemental Material for:\ Localized orbital scaling correction for
       periodic systems}
\author{Aaron Mahler}
\affiliation{Duke University, Department of Physics, Durham, NC 27708}
\author{Jacob Williams}
\affiliation{Duke University, Department of Chemistry, Durham, NC 27708}
\author{Neil Qiang Su}
\affiliation{Department of Chemistry, Key Laboratory of Advanced Energy Materials
             Chemistry (Ministry of Education) and Renewable Energy Conversion
            and Storage Center (RECAST), Nankai University, Tianjin 300071, China
            }
\author{Weitao Yang}
\email{weitao.yang@duke.edu}
\affiliation{Duke University, Department of Chemistry, Durham, NC 27708}
\affiliation{Duke University, Department of Physics, Durham, NC 27708}
\date{\today}

\maketitle



\section{The LOSC Hamiltonian correction}
Recall from the main text that the energy correction due to LOSC is
\begin{equation}
    \Delta E = \frac12 \sum_{ij} \conj{\lambda}_{ij}(\delta_{ij} - \lambda_{ij})
        \wt{\kappa}_{ij},
\end{equation}
where $\conj{\lambda}$ is the complex conjugate of $\lambda$, $\delta_{ij}$ is
the Kronecker delta function, and we have combined the band and unit cell
indices of the dually localized Wannier functions (DLWFs) into a single label
$i$. Then, under the frozen orbital approximation, the LOSC correction to the
Hamiltonian is
\begin{equation}
    \Delta v = \left. \frac{\delta \Delta E}{\delta \rho_s(\bx,\bx')}
        \right\rvert_{\{w_i\}},
\end{equation}
where 
\begin{equation}
\rho_s = \sum_{ij} \lambda_{ij} \outt{w_i}{w_j}
\end{equation}
is the Kohn--Sham density matrix. Note that $\rho_s$ is not diagonal in the
basis of DLWFs; furthermore, unlike the canonical orbital occupations, which
are either 0 or 1, DLWFs can include virtual orbital character, and their
occupations are given by $\lambda_{ij} = \mel{w_i}{\rho_s}{w_j} \in \Comp$.
In the position basis, we write
\begin{equation}
    \rho_s(\bx,\bx') = \sum_{ij} \lambda_{ij} w_i(\bx) \conj{w}_j(\bx'),
\end{equation}
where $\bx = (\br, \sigma)$ combines space and spin variables. Then we may write
the Hamiltonian correction as
\begin{equation}
    \Delta v = \sum_{ij} \left[ 
        \frac{\partial \Delta E}{\partial \lambda_{ij}}
            \frac{\delta \lambda_{ij}}{\delta \rho_s(\bx,\bx')} +
        \frac{\partial \Delta E}{\partial \conj{\lambda}_{ij}}
            \frac{\delta \conj{\lambda}_{ij}}{\delta \rho_s(\bx,\bx')} \right].
\end{equation}

Let us consider the terms in this expression individually. First, consider the
(Wirtinger \cite{wirtinger_zur_1927}) partial derivative of $\Delta E$ with
respect to $\lambda_{ij}$ and its complex conjugate. The diagonal elements
($i = j$) give
\begin{equation}
    \frac{\partial \Delta E}{\partial \lambda_{ii}} = 
        -\frac12 \conj{\lambda}_{ij} \wt{\kappa}_{ii}; \quad
    \frac{\partial \Delta E}{\partial \conj{\lambda}_{ii}} =
        \frac12 (1 - \lambda_{ii}) \wt{\kappa}_{ii}.
\end{equation}
If $i \neq j$, on the other hand,
\begin{equation}
    \frac{\partial \Delta E}{\partial \lambda_{ij}} =
        -\frac12 \conj{\lambda}_{ij} \wt{\kappa}_{ij}; \quad
    \frac{\partial \Delta E}{\partial \conj{\lambda}_{ij}} =
        -\frac12 \lambda_{ij} \wt{\kappa}_{ij}.
\end{equation}
Finally, consider the variation in the local occupations with respect to the
density matrix. From the relationship of $\rho_s$ and $\lambda_{ij}$, we obtain
\begin{equation}
    \frac{\delta \lambda_{ij}}{\delta \rho_s(\bx, \bx')}
        = w_j(\bx) \conj{w}_i(\bx'); \quad
    \frac{\delta \conj{\lambda}_{ij}}{\delta \rho_s(\bx, \bx')}
        = w_i(\bx) \conj{w}_j(\bx').
\end{equation}

Combining yields
\begin{equation}
    \begin{split}
    \Delta v = \frac12 \bigg[ &\sum_i \left(1 - 2\lambda_{ii}\right)
    \wt{\kappa}_{ii} w_i(\bx) \conj{w}_i(\bx') \\
    - &\sum_{i \neq j} \left(
    \lambda_{ij} \wt{\kappa}_{ij} w_j(\bx) \conj{w}_i(\bx') + 
    \conj{\lambda}_{ji} \wt{\kappa}_{ij} w_i(\bx) \conj{w}_j(\bx') \right)
    \bigg].
    \end{split}
\end{equation}

Noting that the matrix $(\lambda_{ij})$ of local occupations is Hermitian and
that the curvature matrix $(\wt{\kappa}_{ij})$ is real symmetric, we can rewrite
our expression in operator form as
\begin{equation}
\begin{split}
    \Delta v 
        &= \sum_{ij} \left(\frac12 \delta_{ij} - \Re \lambda_{ij} \right)
            \wt{\kappa}_{ij} \outt{w_i}{w_j} \\
        &= \frac12 \sum_{ij} \left(\frac12 \delta_{ij} - \lambda_{ij} \right) 
            \wt{\kappa}_{ij} \outt{w_i}{w_j} + \text{h.c.},
\end{split}
\end{equation}
as seen in the main text.

\section{Energy correction with disentanglement}
The construction of dually localized Wannier functions (DLWFs) requires a set of
composite Bloch bands:\ that is, one which is separated by an energy gap from
all other bands at each point in the Brillouin zone. The valence bands of
semiconductors and insulators form a composite set by definition; however,
including virtual bands of semiconductors means that a composite set may not be
attainable, and metals may not have any composite set of bands at all. We refer
to a set of bands which cannot be separated by an energy gap from the rest as
\emph{entangled}. Souza et al.\ showed in \cite{souza_maximally_2001} that
from $N_b$ entangled bands a set of $N_w < N_b$ of Bloch orbitals at each
$\bk$-point can be obtained; this procedure is called \emph{disentanglement}.
Note that these disentangled orbitals are no longer eigenfunctions of the
single-particle Hamiltonian, but they may be treated as a composite set. When
LOSC is applied to the DLWFs found from disentangled Bloch orbitals, the
orbital energy correction is applied to the disentangled Bloch orbitals to
obtain the correction to the frontier orbital energies and the band gap. If
the disentangled set of eigenvalues is a subset of the true eigenvalues in an
area of interest (close to the Fermi energy and the conduction band minimum),
then correcting the disentangled eigenvalues is equivalent to correcting the
original eigenvalues. We found in \cite{mahler_wannier_2022} that including
sufficiently many virtual orbitals in the LOSC procedure satisfies this
condition. For completeness, we demonstrate the form of the eigenvalue
correction when applied to the original Bloch eigenvalues. This can be found by
finding the expectation of the LOSC Hamiltonian correction, $\Delta v$, in the
Bloch orbital basis.

To begin, let us represent the disentangled Bloch orbitals $\ket{\varphi_n\pbk}$ in the original Bloch orbital basis $\ket{\psi_b\pbk}$,
\begin{equation}
    \ket{\varphi_n\pbk} = \sum_b^{N_b} V_{nb}\pbk \ket{\psi_b\pbk}.
\end{equation}
Here, $V\pbk$ is an $N_w$$\times$$N_b$ matrix that obeys the relation $V\pbk(V\pbk)^\dagger = I_{N_w}$, where $I_{N_w}$ is the $N_w$$\times$$N_w$ identity operator. We may then write a transformed Bloch orbital for generalized Wannier function construction in the disentangled Bloch orbital basis as
\begin{equation}
    \ket{\phi_n\pbk} = \sum_a^{N_w} U_{na}\pbk \ket{\varphi_a\pbk}.
\end{equation}
A generalized Wannier function constructed from a set of disentangled Bloch orbitals then takes the form 
\begin{equation}
\begin{split}
    \ket{w_m^{\bR}} &= \frac{1}{N_k} \sum_{\bk}^{N_k} e^{-i\bk\cdot\bR}
    \sum_a^{N_w} U_{ma}\pbk \sum_b^{N_b} V_{ab}\pbk \ket{\psi_b\pbk} \\
    &= \frac{1}{N_k} \sum_{\bk}^{N_k} e^{-i\bk\cdot\bR}  
    \sum_b^{N_b} L_{mb}\pbk \ket{\psi_b\pbk},
\end{split}
\end{equation}
where $L_{mb}\pbk = [U\pbk V\pbk]_{mb}$. 
Recall that the LOSC method corrects the Kohn--Sham Hamiltonian by
\begin{equation}
\begin{split}
    \Delta v 
        &= \frac{\delta \Delta E^{\text{LOSC}}}{\delta \rho_s} \\
        &= \sum_{\bR mn} \widetilde{\kappa}_{mn}^{\bZ\bR} \text{Re}\left\{
    \left(\frac{\delta_{mn}^{\bZ\bR}}{2} - \lambda_{mn}^{\bZ\bR} \right) 
    \outt{ w_m^{\bZ} }{ w_n^{\bR} } \right\}.
\end{split}
\end{equation}
To compute the matrix elements of $\Delta v$ in the basis of Bloch orbitals, we require the overlap of a Wannier function and Bloch orbital:
\begin{equation}
\begin{split}
    \inn{w_n^\bR}{\psi_i^\bk} &= 
    \frac{1}{N_k} \sum_\bq e^{i\bq\cdot\bR} \inn{\phi_n^\bq}{\psi_i^\bk} \\
    &= \frac{1}{N_k} \sum_\bq e^{i\bq\cdot\bR} 
    \sum_b^{N_b} \overline{L}_{nb}^\bq \inn{\psi_b^\bq}{\psi_i^\bk} \\
    &= e^{i\bk\cdot\bR} \overline{L}_{ni}^\bk,
\end{split}
\end{equation}
where we have used the normalization convention $\inn{\psi_b^\bq}{\psi_i^\bk}=N\delta_{\bq\bk}\delta_{bi}$. We write $\overline{z}$ for the complex conjugate of $z$. We can now easily find the expectation of the LOSC Hamiltonian correction in the Bloch orbital basis,
\begin{equation}
\begin{split}
    \mel{\psi_i\pbk}{\Delta v}{\psi_i\pbk}
    &= \sum_{\bR mn} \widetilde{\kappa}_{mn}^{\bZ\bR} \text{Re}\left\{
    \left(\frac{\delta_{mn}^{\bZ\bR}}{2} - \lambda_{mn}^{\bZ\bR} \right) 
    \inn{\psi_i^\bk}{ w_m^{\bZ} }
    \inn{ w_n^{\bR} }{\psi_i^\bk} \right\} \\
    &=  \sum_{\bR mn} \widetilde{\kappa}_{mn}^{\bZ\bR} \text{Re}\left\{
    \left(\frac{\delta_{mn}^{\bZ\bR}}{2} - \lambda_{mn}^{\bZ\bR} \right)
    L_{mi}^\bk \overline{L}_{ni}^\bk e^{i\bk\cdot\bR}.
    \right\}
\end{split}
\end{equation}
Functionally, the only difference between this form and using the disentangled basis is that instead of $L^\bk$ the correction would have $U^\bk$, the transformation from the (possibly) disentangled basis to the transformed basis.

\section{The Coulomb integral}
In this section, we discuss details of the Coulomb interaction $J$ between two
dually localized Wannier functions (DLWFs) $\ket{w_i}$ and $\ket{w_j}$; note
that we combine the DLWFs' band-like and unit cell indices into a single
subscript, so $\ket{w_i} = \ket{w_m^{\bR}}$. $J$ depends on the charge density
$\rho$ associated with each DLWF, $\rho_i(\br) = |w_i(\br)|^2$, and is given by
\begin{equation}
    J[\rho_i,\rho_j] 
        = \iint d\br\, d\br'\, \frac{\rho_i(\br)\rho_j(\br')}{|\br-\br'|}
        = \iint d\br\, d\br'\, \rho_i(\br)\rho_j(\br')\, K(|\br-\br'|),
\end{equation}
where we term $K(r) = 1/r$ the \emph{Coulomb kernel}.

Na\"ively, the cost of computing $J[\rho_i,\rho_j]$ scales as $O(N_{\br})^2$,
where $N_{\br}$ is the size of the densities on the real grid. However, the
DLWF densities are periodic on the Born--von Karman supercell (see the main
text). Recall that we may write the Fourier series for sufficiently well-behaved
(that is, obeying Dirichlet conditions) periodic functions $f(\br)$ as
\begin{equation}
    f(\br) = \frac{1}{\Omega_{\text{BvK}}}\sum_{\bG} f(\bG) e^{i\bG \cdot \br},
\end{equation}
where the basis vectors $\bG$ belong to the reciprocal lattice and
$\Omega_{\text{BvK}}$ is the volume of the Born--von Karman supercell. Each
DLWF density can be so represented, yielding $N_{\bG}$  Fourier coefficients
$\rho(\bG)$. We can then exploit the fast Fourier transform  to compute
$J[\rho_i,\rho_j]$ in $O(N_{\bG} \log N_{\bG})$ time.

The Fourier-space representation of $J[\rho_i,\rho_j]$ is a classical result
in electrodynamics, but we present it briefly here. First, observe that we
can write
\begin{equation}
    J[\rho_i,\rho_j] 
        = \iint d\br\, d\br'\, \frac{\rho_i(\br)\rho_j(\br')}{|\br-\br'|}
        = \int d\br\, V_i(\br) \rho_j(\br),
\end{equation}
where $V_i$ is the potential due to $\rho_i$ experienced by $\rho_j$,
\begin{equation}
    V_i(\br) = \int d\br'\, \frac{\rho_i(\br')}{|\br-\br'|}.
\end{equation}
$V_i$ inherits the periodicity of $\rho_i$ and can be written as a Fourier
series; it also satisfies the Poisson equation
\begin{equation}
    \nabla^2 V_i(\br) = -4\pi \rho_i(\br).
\end{equation}
This equation is algebraic in the plane-wave basis since
$\nabla^2 e^{i\bG\cdot\br} = -G^2$, where $G = \abs{\bG}$; thus
\begin{equation}
    \nabla^2 \frac{1}{\Omega_{\text{BvK}}}
            \left( \sum_{\bG} V_i(\bG) e^{i\bG\cdot\br} \right)
        = -\frac{1}{\Omega_{\text{BvK}}}
            \sum_{\bG} G^2 V_i(\bG) e^{i\bG\cdot\br}
        = -\frac{4\pi}{\Omega_{\text{BvK}}}
            \sum_{\bG} \rho_i(\bG) e^{i\bG\cdot\br}.
\end{equation}
Since the plane waves $e^{i\bG\cdot\br}$ are orthonormal, we can solve this
equation term by term, obtaining the Fourier coefficients
\begin{equation}
    V_i(\bG) = 4\pi \frac{\rho_i(\bG)}{G^2}.
\end{equation}
Substituting $V_i$ into the original expression and replacing $\rho_j$ by
its Fourier series gives
\begin{equation}
\begin{split}
    J[\rho_i,\rho_j] 
        &= \frac{4\pi}{\Omega_{\text{BvK}}^2} \int d\br\, 
            \left(\sum_{\bG} \frac{\rho_i(\bG)}{G^2} 
                e^{i\bG\cdot\br} \right)
            \left(\sum_{\bG'} \rho_j(\bG') e^{i\bG'\cdot\br} \right) \\
        &= \frac{4\pi}{\Omega_{\text{BvK}}^2} \int d\br\,
            \left(\sum_{\bG} \frac{\conj{\rho}_i(\bG)}{G^2} 
                e^{-i\bG\cdot\br} \right)
            \left(\sum_{\bG'} \rho_j(\bG') e^{i\bG'\cdot\br} \right) \\
        &= \frac{4\pi}{\Omega_{\text{BvK}}^2} \int d\br\, \sum_{\bG\bG'} 
            \frac{\conj{\rho}_i(\bG)\rho_j(\bG')}{G^2} 
            e^{i(\bG'-\bG)\cdot\br}
         = \frac{4\pi}{\Omega_{\text{BvK}}}
            \sum_{\bG} \frac{\conj{\rho}_i(\bG)\rho_j(\bG)}{G^2}.
\end{split}
\end{equation}
We have that $\rho_i(-\bG) = \conj{\rho}_i(\bG)$, where $\conj{\rho}$ is the
complex conjugate of $\rho$, since $\rho_i(\br)$ is real. The integral and
double sum over $\bG,\bG'$ collapse into a single sum by the orthonormality
of the plane waves $e^{i\bG\cdot\br}$:
\begin{equation}
    \int d\br\, e^{i(\bG-\bG')\cdot\br} = \Omega_{\text{BvK}} \delta_{\bG,\bG'},
\end{equation}
where $\delta_{\bG,\bG'}$ is the Kronecker delta function.

In perfect analogy with the real-space description, we can express the Coulomb integral in reciprocal space as a sum over Fourier coefficients with a reciprocal-space Coulomb kernel. That is,
\begin{equation}
    J[\rho_i,\rho_j] = \frac{1}{\Omega_{\text{BvK}}} \sum_{\bG} \conj{\rho_i}(\bG)\rho_j(\bG)K(G),
\end{equation}
where $K(G) = 4\pi/G^2$. This kernel has a singularity at $G=0$, as does
$K(r)$; however, unlike its real-space counterpart, the singularity in $K(G)$ is
not integrable in three dimensions.

For the density of the entire unit cell, the singularity is circumvented by
requiring the cell to be charge neutral, which is equivalent to setting $K(0) = 0$.
However, the DLWFs are individually charged, so an alternative approach must be
chosen. We address two problems in this work:\ the spurious interaction of a
DLWF density with its periodically repeating images, and the suppression by the
bulk environment of the long-range behavior of the Coulomb kernel. Both of these
can be addressed by modifying the Coulomb kernel $K(r)$.

\subsection{The spherical cutoff kernel}
The Fourier coefficients $K(G)$ of the Coulomb kernel can be derived by taking
the Fourier transform of the Yukawa kernel $K_y(r) = e^{-\lambda r}/r$ in the
limit $\lambda \to 0$. Likewise, the Fourier transforms of other modified
kernels $K_{\text{mod}}(r)$ yield the modified Coulomb interaction in the
plane-wave basis as
\begin{equation}
    J_{\text{mod}}[\rho_i,\rho_j]
        = \iint d\br\,d\br'\,\rho_i(\br)\rho_j(\br')K_{\text{mod}}(|\br-\br'|)
        = \frac{1}{\Omega_{\text{BvK}}} \sum_{\bG}
            \conj{\rho}_i(\bG)\rho_j(\bG)K_{\text{mod}}(|\bG|).
\end{equation}
In order to prevent spurious interaction with the densities' periodic images,
we use the spherical cutoff kernel $K_c$, used first by Onida and coworkers in
\cite{onida_ab_1995} for $GW$ calculations and applied to DFT by Jarvis and
coworkers in \cite{jarvis_supercell_1997}. It is defined as 
\begin{equation}
    K_c(r; R_c) = \frac{1 - \Theta(r - R_c)}{r} =
    \begin{cases}
        \dfrac1r & r < R_c \\
        0        & r \geq R_c,
    \end{cases}
\end{equation}
where $\Theta(r)$ is the Heaviside step function. $R_c$ is the cutoff radius,
taken here to be half the length of the shortest Born--von Karman supercell
lattice vector. Performing the Fourier transform yields
\begin{equation}
    K_c(G; R_c) = 
    \begin{cases}
        \dfrac{4\pi}{G^2} \left[1 - \cos(GR_c)\right] & G \neq 0 \\
        2\pi R_c^2 & G = 0.
    \end{cases}
\end{equation}
Note that there is no singularity in $K_c(G; R_c)$. The sinusoidal oscillations are
expected; they exemplify the Gibbs phenomenon due to the jump discontinuity of
$K_c(r; R_c)$ at $R_c$.

\subsection{The sLOSC kernel}
Even after introducing the spherical cutoff, the effect of the lattice means
that the long-range $1/r$ behavior of $J[\rho_i,\rho_j]$ must be attenuated to
give an accurate correction to the band gap of bulk systems. To complete the
sLOSC kernel, we therefore introduce an additional screening term proportional
to the complementary error function $\erfc(r) = 1 - \erf(r)$, yielding
\begin{equation}
    K_s(r; R_c, \alpha)
        = \frac{\erfc(\alpha r)\left[1 - \Theta(r - R_c)\right]}{r}
        = \begin{cases}
            \dfrac{\erfc(\alpha r)}{r} & r < R_c \\
            0 & r \geq R_c.
          \end{cases}
\end{equation}
Let us derive the reciprocal-space kernel $K_s(G)$ by Fourier transforming
$K_s(r)$ directly, utilizing the rotational symmetry of the real-space kernel
to convert into spherical coordinates. Thus
\begin{equation}
\begin{split}
    K_s(G; R_c, \alpha) 
        &= \int d\br\, K_s(r; R_c, \alpha) e^{-i \bG \cdot \br} \\
        &= \int_0^{2\pi} d\theta\, \int_0^\pi d\phi\, 
            \int_0^\infty dr\, r^2 \sin \phi
                \frac{\erfc(\alpha r)\left[1 - \Theta(r - R_c)\right]}{r} 
                e^{-i G r \cos \phi} \\
        &= 2\pi \int_0^\pi d\phi\, \sin \phi
            \int_0^{R_c} r \erfc(\alpha r) e^{-i G r \cos \phi}.
\end{split}
\end{equation}
Setting $u = \cos \phi$, whence $du = -d\phi\, \sin \phi$, we next obtain
\begin{equation}
\begin{split}
    K_s(G; R_c, \alpha) &= 2\pi \int_0^{R_c} d\br\, r \erfc(\alpha r)
        \int_{-1}^1 du\, e^{-iGru}
         = 2\pi \int_0^{R_c} d\br\, r \erfc(\alpha r) 
            \left[\frac{e^{iGr} - e^{-iGr}}{iGr}\right] \\
        &= \frac{4\pi}{G} \int_0^{R_c} d\br\, \erfc(\alpha r) \sin(Gr).
\end{split}
\end{equation}
Performing the final integral with the Mathematica software package
\cite{wolfram_research_inc_mathematica_2021} yields
\begin{multline}
    K_s(G; R_c, \alpha) = \frac{4\pi}{G^2} \Bigg\{ 1 - \cos(GR_c) \erfc(\alpha R_c) + \\
        \frac{e^{-G^2/4\alpha^2}}{2} 
        \left[ \erf\left(- \alpha R_c + \frac{iG}{2\alpha}\right) 
        - \erf\left(\alpha R_c + \frac{iG}{2\alpha}\right) \right] \Bigg\}.
\end{multline}
Writing $z = \alpha R_c + iG/2\alpha$ and noting that
$\erf(\conj{z}) = \conj{\erf z}$ and $\erf(-z) = -\erf z$, we may simplify the
relevant part of $K_s(G; R_c, \alpha)$:
\begin{equation}
    \erf(-\conj{z}) - \erf(z) = - (\erf{\conj{z}} + \erf z)
        = - (\conj{\erf z} + \erf z) = -2 \Re\{ \erf z \}.
\end{equation}
Thus, we obtain in accordance with the main text that when $G \neq 0$
\begin{equation}
    K_s(G; R_c, \alpha) = \frac{4\pi}{G^2} \left[ 1 - \cos(GR_c) 
        \erfc(\alpha R_c) - e^{-G^2/4\alpha^2} \Re \left\{ \erf 
        \left(\alpha R_c + \frac{iG}{2\alpha} \right) \right\} \right].
\end{equation}
Note that $K_s(G)$ is not singular at $G = 0$; we have
\begin{equation} \label{eq:codekerzero}
    K_s(G = 0; \alpha) = 2\pi R_c^2 
        + \pi \erf (\alpha R_c) \left(\frac{1}{\alpha^2} - 2 R_c^2\right)
        - \frac{2\sqrt{\pi}R_c e^{-(\alpha R_c)^2}}{\alpha}.
\end{equation}

\subsubsection{The Faddeeva function}
By Liouville's theorem, $\erf z$ is unbounded for $z \in \Comp$; in practice,
it overflows floating-point arithmetic even for $\bG$-vectors of relatively
small modulus. Therefore, to use the sLOSC kernel in practice, we require a 
scaled implementation of the complex error function, namely the Faddeeva
function
\begin{equation}
    w(z) = e^{-z^2} \erfc(-iz).
\end{equation}
In particular, we seek to replace
\begin{equation*}
    e^{-G^2/4\alpha^2} \Re \left\{ \erf 
        \left(\alpha R_c + \frac{iG}{2\alpha} \right) \right\}
\end{equation*}
by a term containing $w(z)$. As mentioned in the main text, we implement the
Faddeeva function using ACM Algorithm 916
\cite{zaghloul_algorithm_2012,zaghloul_remark_2016}.

Note that whenever $\Re z > 0$ we may write
\begin{equation}
    \erf z = 1 - e^{-z^2} w(iz);
\end{equation}
in our case, $\Re z = \alpha R_c > 0$. Thus, we may write
\begin{equation}
\begin{split}
    e^{-G^2/4\alpha^2} &\Re \left\{ \erf 
            \left(\alpha R_c + \frac{iG}{2\alpha} \right) \right\} \\
        &= e^{-G^2/4\alpha^2} \Re \left\{ 1 - e^{-(\alpha R_c + iG/2\alpha)^2}\,
            w\left(i\alpha R_c - \frac{G}{2\alpha}\right) \right\} \\
        &= e^{-G^2/4\alpha^2} - \Re \left\{ 
            e^{-(\alpha R_c + iG/2\alpha)^2 - (G/2\alpha)^2}\,
            w\left(i\alpha R_c - \frac{G}{2\alpha}\right) \right\} \\
        &= e^{-G^2/4\alpha^2} - \Re \left\{
            e^{-(\alpha R_c)^2} e^{-iGR_c}\,
            w\left(i\alpha R_c - \frac{G}{2\alpha}\right) \right\} \\
        &= e^{-G^2/4\alpha^2} - e^{-(\alpha R_c)^2} \Re \left\{ 
            e^{-i G R_c} \,
            w\left(i\alpha R_c - \frac{G}{2\alpha}\right) \right\}.
\end{split}
\end{equation}
Breaking the Faddeeva function into its real and imaginary parts as
\begin{equation}
   w\left(i\alpha R_c - \frac{G}{2\alpha}\right) = V + iL,
\end{equation}
we obtain
\begin{equation}
\begin{split}
    \Re \left\{ e^{-i G R_c}\,
        w\left(i\alpha R_c - \frac{G}{2\alpha}\right) \right\} =
    \Re &\left\{ \left[\cos (GR_c) - i \sin (GR_c) \right]\,
        w\left(i\alpha R_c - \frac{G}{2\alpha}\right) \right\} \\
    &= \Re \left\{ \left[\cos (GR_c) - i \sin (GR_c) \right](V + iL) \right\} \\
    &= V\, \cos(GR_c) + L\, \sin(GR_c).
\end{split}
\end{equation}
The Fourier coefficients of the sLOSC kernel as implemented in our code are thus
\begin{multline} \label{eq:codekernel}
    K_s(G; R_c, \alpha) 
        = \frac{4\pi}{G^2} \Bigg\{ 1 - \cos(GR_c) \erfc(\alpha R_c)
            - e^{-G^2/4\alpha^2} + e^{-(\alpha R_c)^2} \times \\
        \Bigg[
        \Re \left\{ w\left(i\alpha R_c - \frac{G}{2\alpha}\right) \right\} \cos(GR_c) +
        \Im \left\{ w\left(i\alpha R_c - \frac{G}{2\alpha}\right) \right\} \sin(GR_c)
        \Bigg] \Bigg\};
\end{multline}
$K_s(G=0)$, having only a real argument to the error function, is implemented as
written in Eq.\ (\ref{eq:codekerzero}).

\section{Computational details}
In any DFT calculation in periodic boundary conditions, several computational
parameters must be selected. In this work, we use the PBE density functional
\cite{perdew_generalized_1996}; set the kinetic energy cutoff of the plane-wave
basis at \SI{100}{\rydberg}; sample the Brillouin zone uniformly by \kmt{16}
$\bk$-points for self-consistent calculations (yielding the Kohn--Sham density
matrix), and \kmt{6} $\bk$-points for the other calculations (conduction Bloch
bands, localization, and sLOSC); and use the norm-conserving optimized
Vanderbilt pseudopotentials with scalar relativistic correction 
\cite{hamann_optimized_2013}, sourced from the PseudoDojo 
\cite{van_setten_pseudodojo_2018}.

Localization is performed using a fork of the \texttt{wannier90} code
\cite{mostofi_wannier90_2008, mostofi_updated_2014, pizzi_wannier90_2020}
maintained by one of us (A.M.) \cite{mahler_wannier_2022}. We set the frozen
(inner) window for disentanglement at \SI{0.5}{\electronvolt} above the Fermi energy. For an initial guess, we use the selected columns of the density matrix (SCDM)\cite{damle_compressed_2015,damle_scdm-k_2017}, with midpoint $\mu$ set at
the top of the frozen window and spread
$\sigma = \SI{4.0}{\electronvolt}$. Worth noting is that some
systems presented difficulty in converging to the minimal value of the cost
function; we are not guaranteed a unique global minimum of our cost function,
and the DLWFs can display imaginary components, especially in the virtual space
\cite{brouder_exponential_2007}. We searched manually over a range of conjugate
gradient step sizes to seek convergence, choosing the one that yielded the
smallest total cost.

Several additional parameters must be set for sLOSC computations. First, the
mixing parameter $\gamma$ in the cost function, which controls the
relative importance of spatial and energy localization of the DLWFs, must be
chosen. In the main text, we use $\gamma = 0.47714$ to match the molecular LOSC
results; see below for a discussion of the effects of varying $\gamma$. The
exchange factor $\tau$ in the sLOSC curvature can also be modified; we choose the
nonempirical value $\tau=6(1-2^{-1/3})$, which enforces the condition
$X[\rho_1] = 2X[\frac12 \rho_1]$ for any one-electron density $\rho_1$ (for a
derivation of this fact, see the Supporting Information of
\cite{li_localized_2018}).

\subsection{The number of virtual orbitals}
Because the DLWFs mix both occupied (valence) and unoccupied (conduction)
states, it is necessary to include enough virtual bands in their construction
that the Fermi level is reproduced accurately. To ensure the frontier orbitals
are converged in the number of virtual states, we chose ten representative
systems and perform sLOSC on them with $n_v = 2, 3, 4, 5, 6$ coordination shells
of virtual Bloch bands. By "coordination shell``, we mean the coordination
number of the lattice; thus, for diamond, zincblende, and wurtzite structures we
include $4 n_v$ virtual , while for cubic structures we include $6 n_v$ virtual
bands. In all cases, we disentangle to one fewer coordination shell of
transformed Bloch orbitals (and, hence, DLWFs). 

In \ref{fig:virtsweep}, we compare the fundamental gap calculated with sLOSC as a
function of the Coulombic screening parameter $\alpha$ (for details on $\alpha$,
see the next subsection), for different numbers of virtual shells. We find that
disentangling to two virtual shells is sufficient to converge the gap; the notable
differences between different numbers of virtual shells come from qualitative
differences in the localization procedure.

\begin{figure}[ht!]
    \centering
    \subfloat[C (diamond)]{\includegraphics[width=0.45\linewidth]{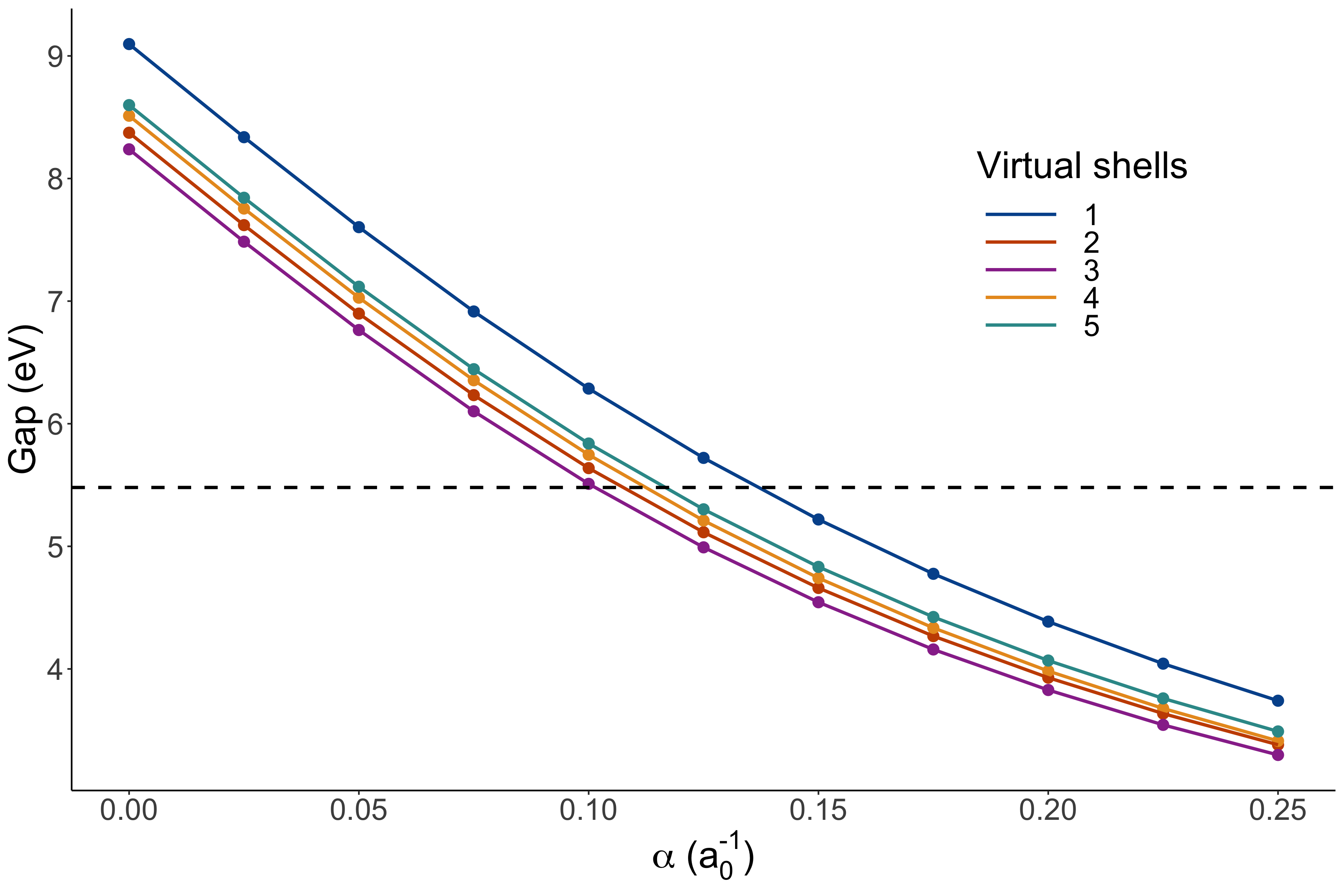}}\quad
    \subfloat[GaSb (zincblende)]{\includegraphics[width=0.45\linewidth]{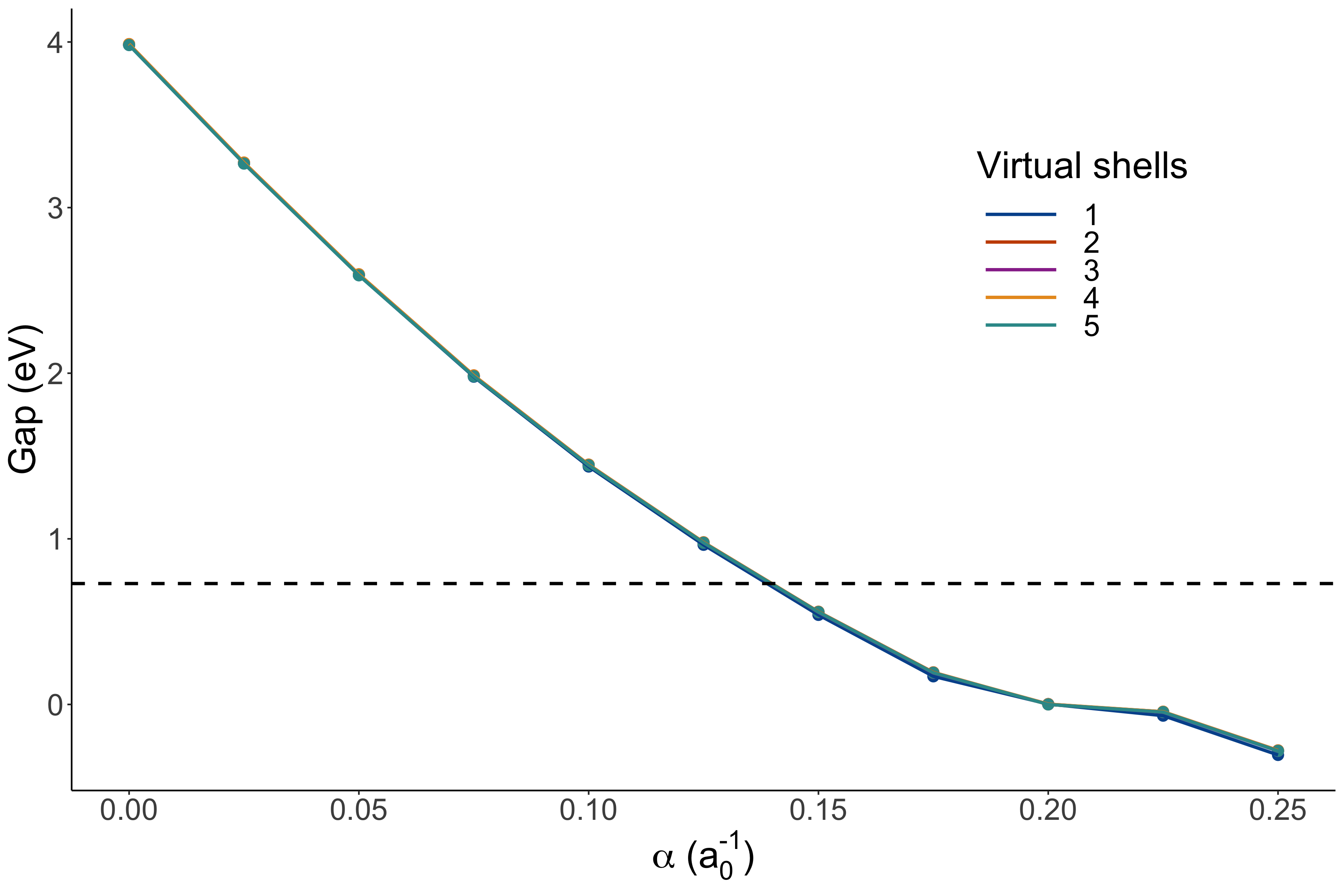}}\\
    \subfloat[Ge (diamond)]{\includegraphics[width=0.45\linewidth]{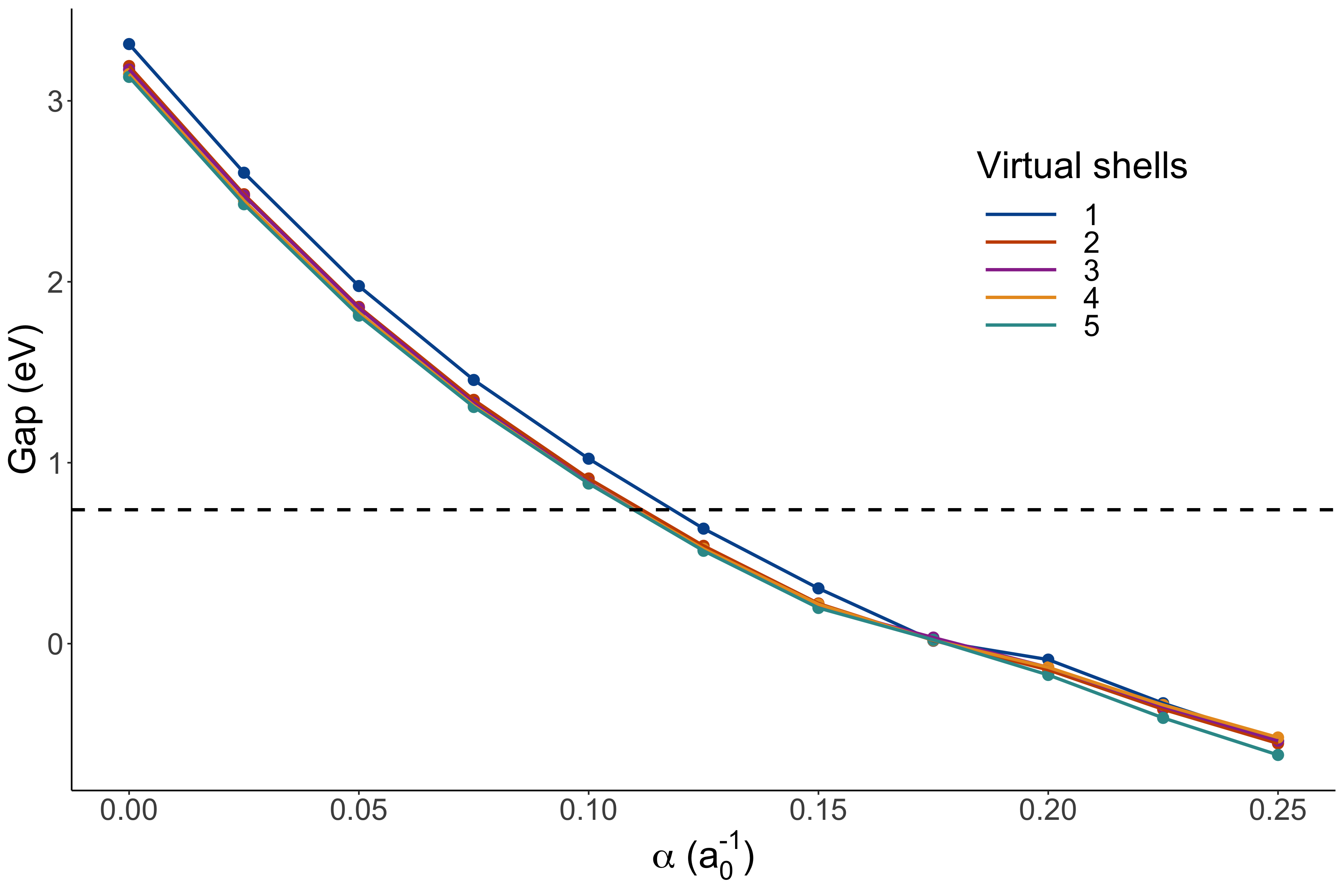}}\quad
    \subfloat[InAs (zincblende)]{\includegraphics[width=0.45\linewidth]{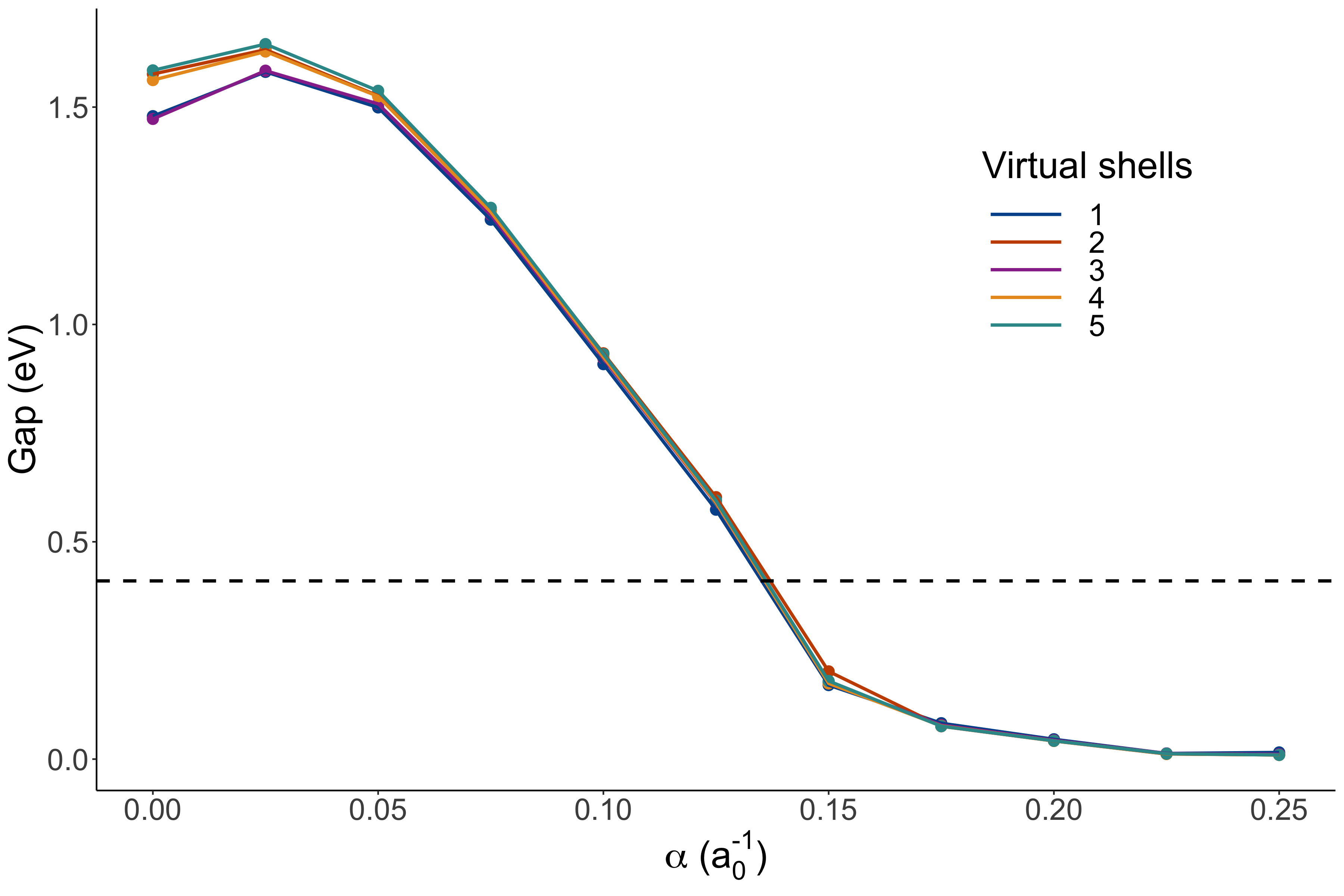}}\\
    \subfloat[InN (wurtzite)]{\includegraphics[width=0.45\linewidth]{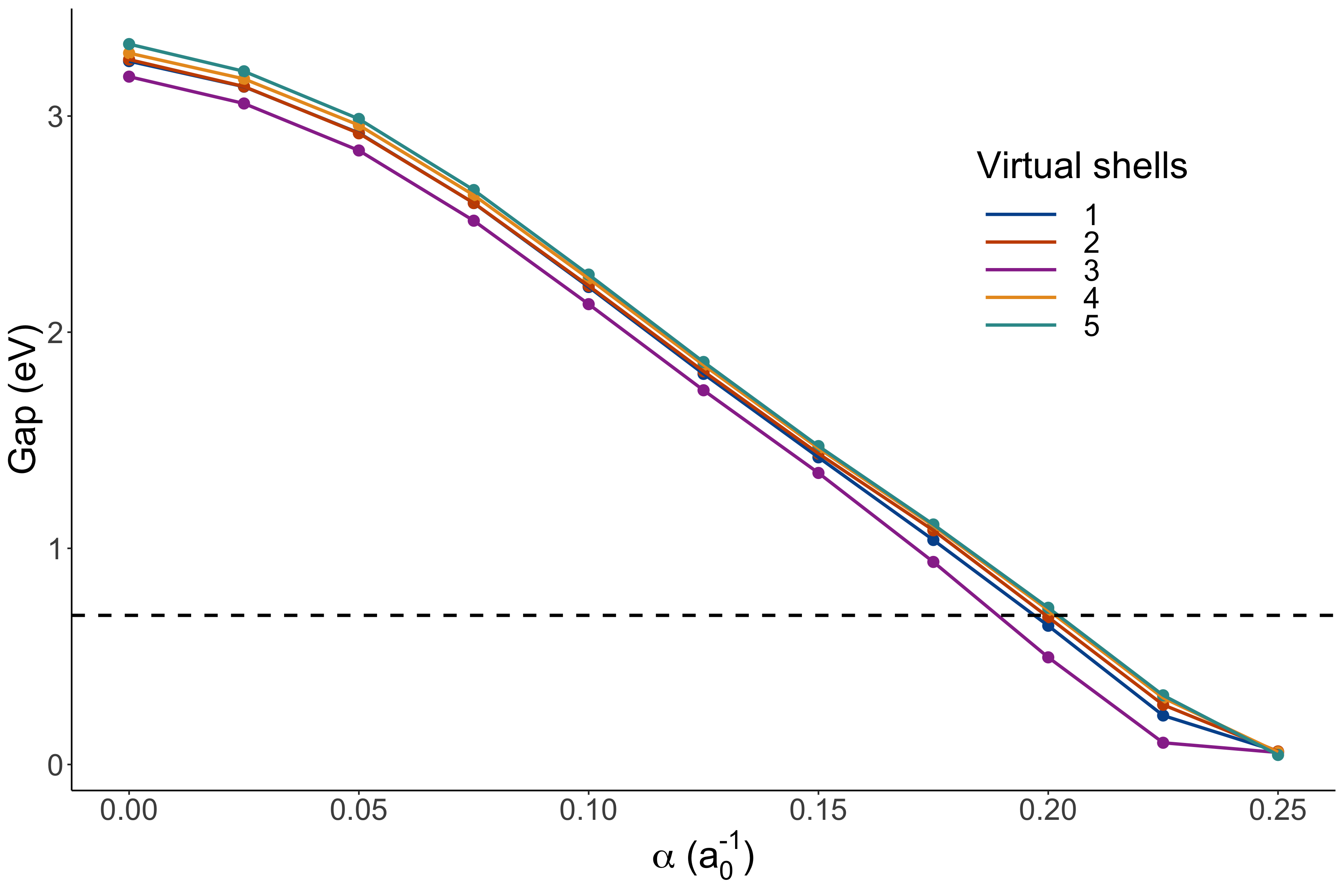}}\quad
    \subfloat[InSb (zincblende)]{\includegraphics[width=0.45\linewidth]{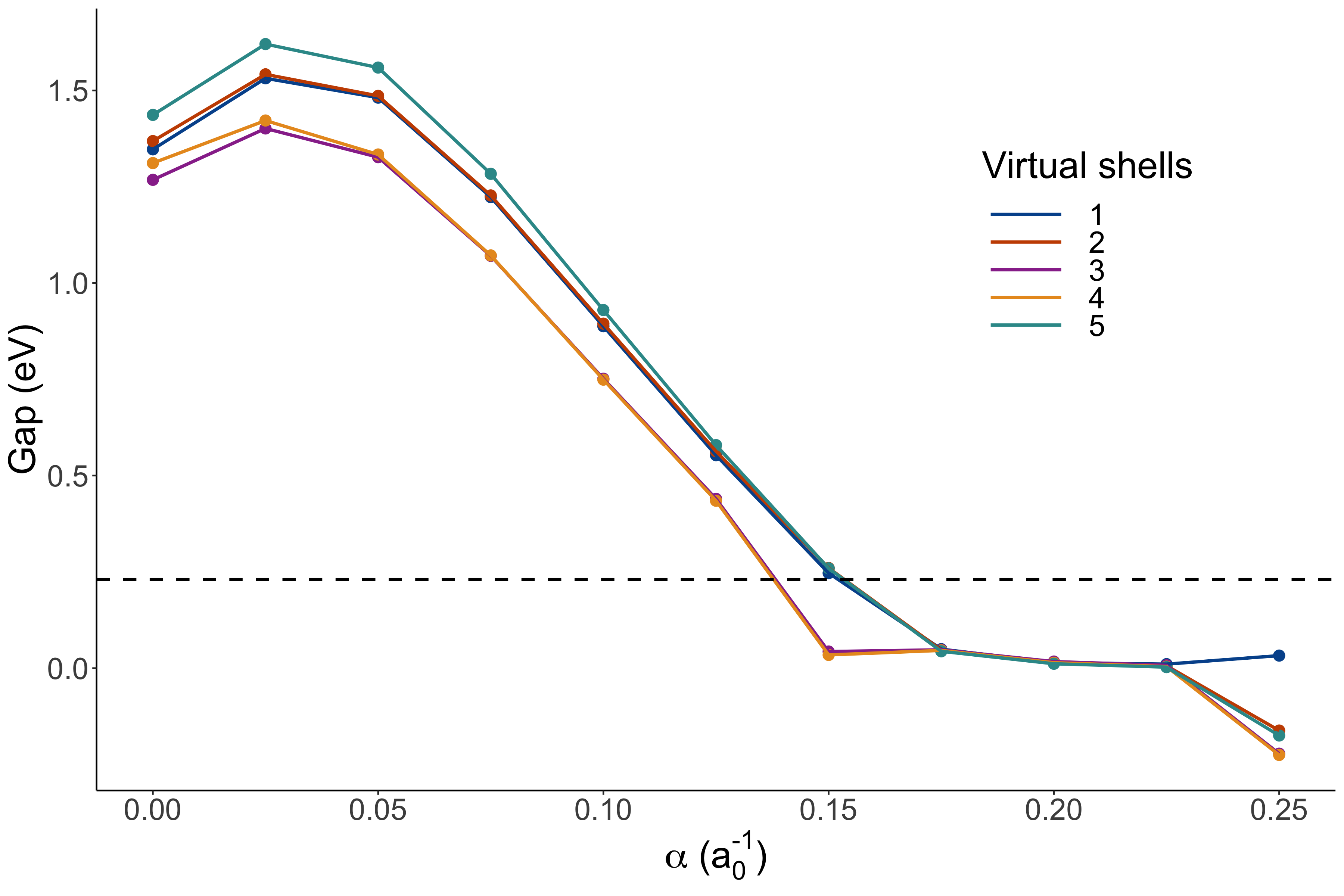}}\\
    \caption{sLOSC band gaps as a function of screening parameter $\alpha$ for
             $n=1,2,3,4,5$ virtual shells of DLWFs. The dashed line is the
             experimental gap. The subcaption title gives the chemical
             formula, with the lattice type in parentheses.}
    \label{fig:virtsweep}
\end{figure}
\begin{figure}[ht!]
    \ContinuedFloat
    \centering
    \subfloat[LiF (cubic)]{\includegraphics[width=0.45\linewidth]{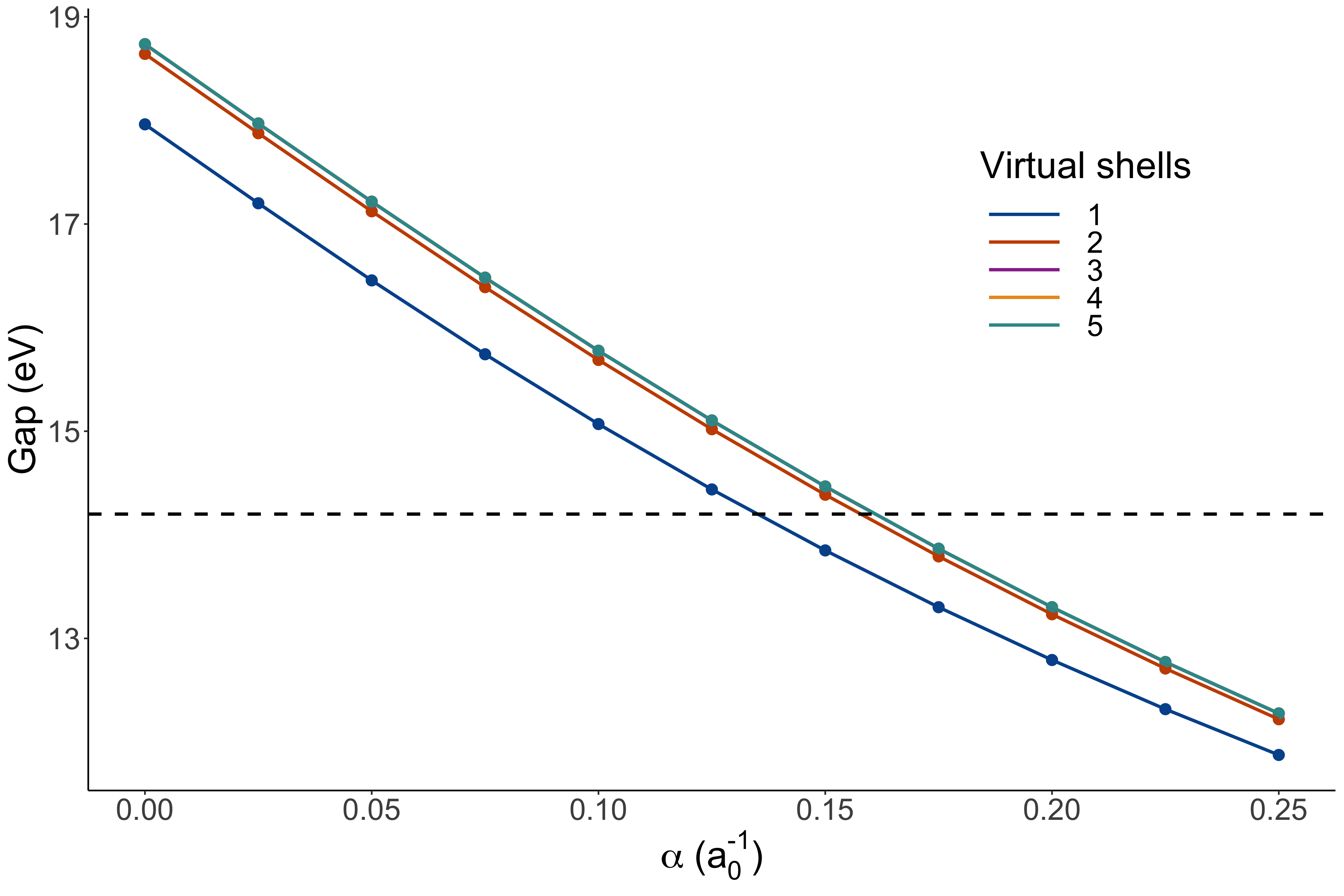}}\quad
    \subfloat[NaF (cubic)]{\includegraphics[width=0.45\linewidth]{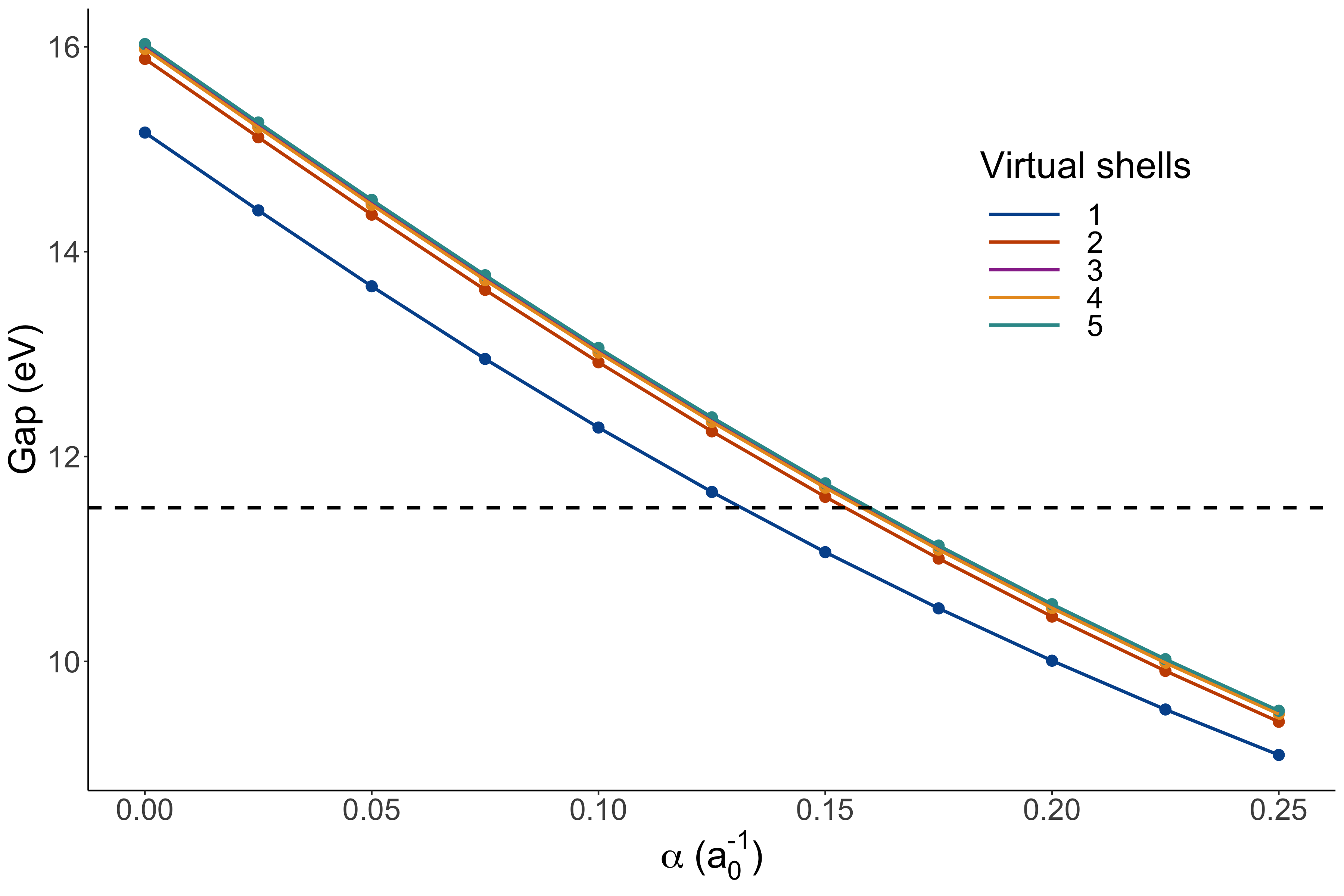}}\\
    \subfloat[Si (diamond)]{\includegraphics[width=0.45\linewidth]{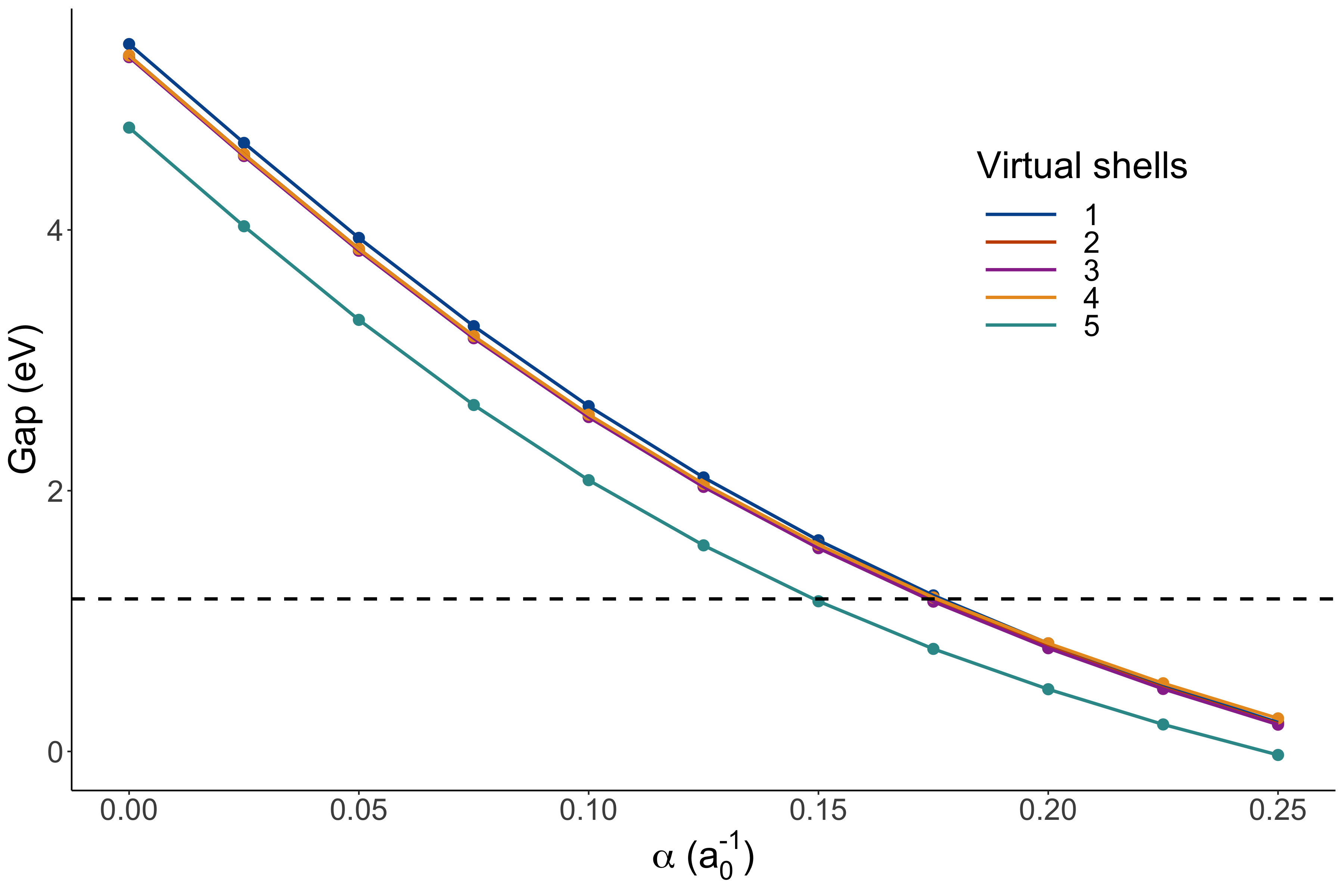}}\quad
    \subfloat[SiC (zincblende)]{\includegraphics[width=0.45\linewidth]{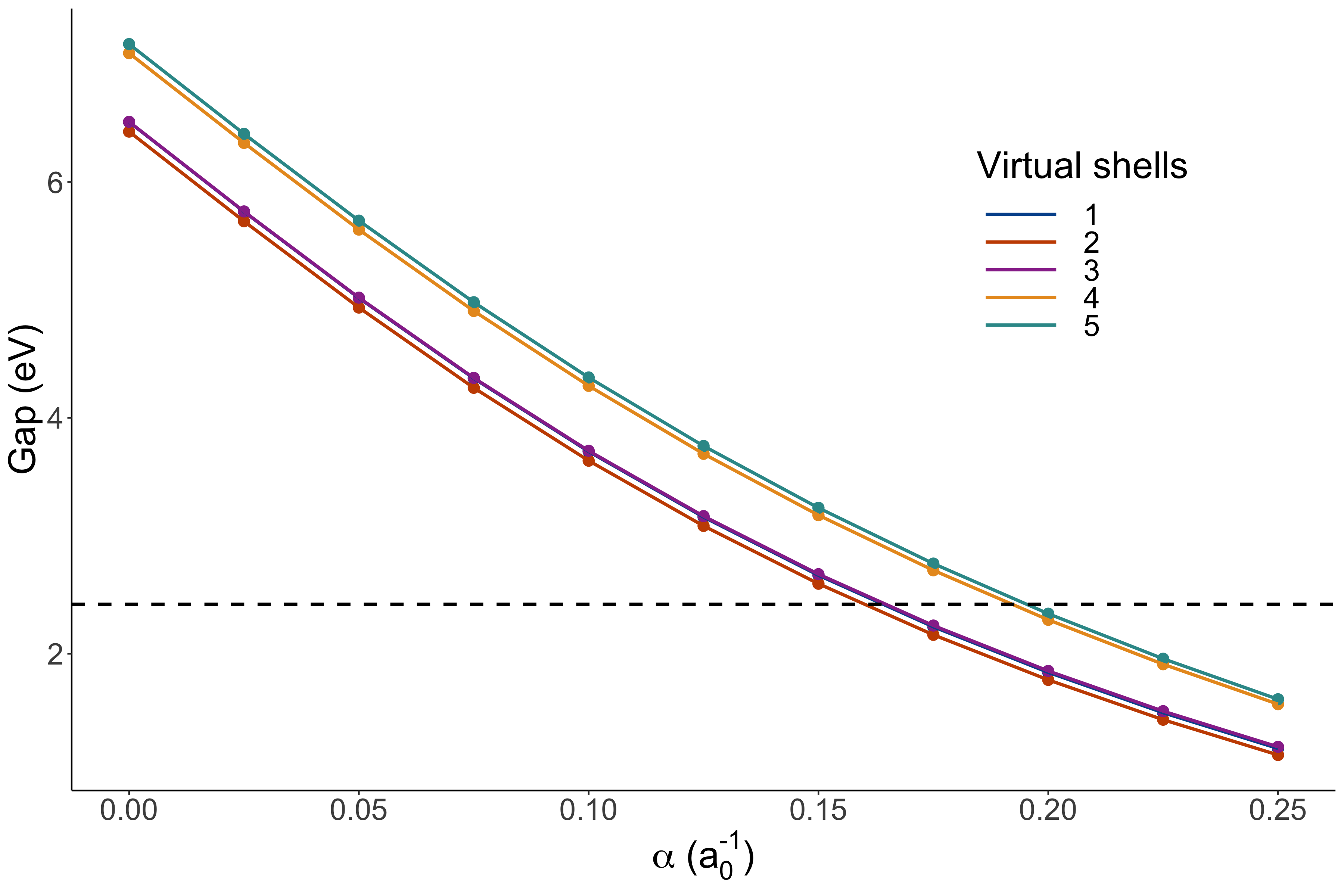}}\\
    \caption{sLOSC band gaps as a function of screening parameter $\alpha$ for
             $n=1,2,3,4,5$ virtual shells of DLWFs. The dashed line is the
             experimental gap. The subcaption title gives the chemical
             formula, with the lattice type in parentheses.}
\end{figure}

\clearpage

\subsection{Choosing the screening parameter}
\begin{figure}[!ht]
    \centering
    \includegraphics[width=0.8\linewidth]{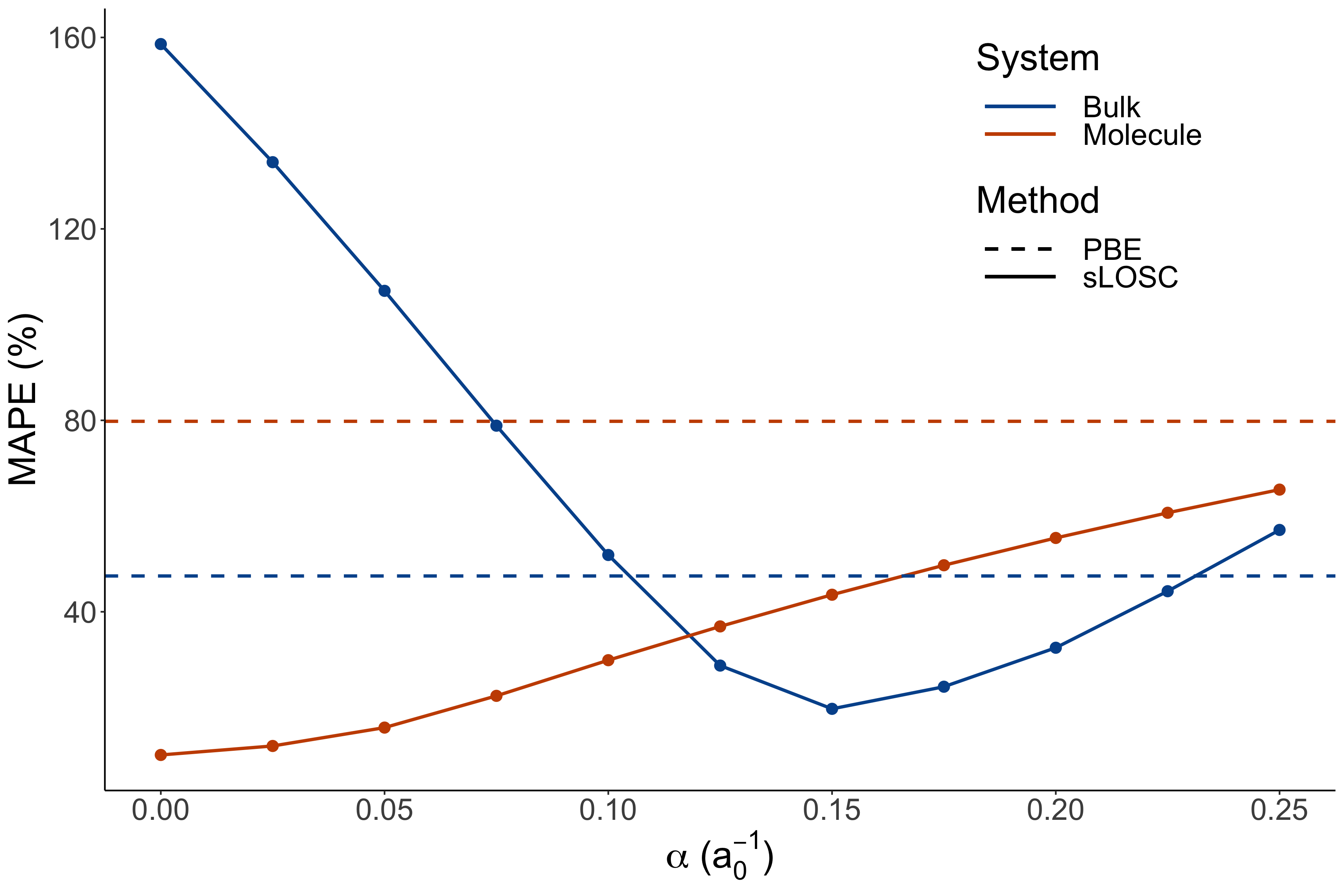}
    \caption{MAPE of the band gap of bulk (orange) and molecular (blue) systems,
             as $\alpha$ is varied. The dashed horizontal lines give the MAPE of
             the uncorrected DFA (PBE).}
    \label{fig:screensweep}
\end{figure}

The sLOSC method's utility in calculating the band gap of bulk materials comes
at the cost of one additional empirical parameter $\alpha$, which measures the
strength of the screening. To choose $\alpha$, we computed the band gaps of the
full dataset below, varying $\alpha$ from 0 (unscreened) to
\SI{0.25}{\bohr^{-1}} in steps of 0.025, where \si{\bohr} denotes the Bohr
radius. We selected the value minimizing the mean absolute percentage error
(MAPE) of the bulk systems' band gaps relative to experimentally determined values. As can be seen from Fig.\ \ref{fig:screensweep}, the minimal MAPE for
bulk systems is attained at $\alpha = \SI{0.15}{\bohr^{-1}}$, so we choose this
value for the screening in sLOSC.

We also show the MAPE for a test set of small molecules' fundamental gaps
(ionization potential minus electron affinity), relative to CCSD(T)
calculations. The smallest MAPE is given by LOSC2 \cite{su_preserving_2020},
with $\alpha = 0$ (no screening); however, even with
$\alpha = $ \SI{0.15}{\bohr^{-1}} as in sLOSC, the MAPE is substantially improved
compared to the uncorrected PBE computation.

For the results of individual uncorrected and sLOSC computations, see
Tables \ref{tab:sc40-data} and \ref{tab:mol-data} in the section below.

\subsection{The DLWF space/energy mixing parameter}
Recall that the DLWFs are obtained by minimizing the cost function
\begin{equation}
    F = (1-\gamma)\sum_i \evo{\Delta r^2}_i + 
        \gamma C \sum_i \evo{\Delta h_\text{s}^2}_i,
\end{equation}
where $\evo{\Delta r^2}_i$ ($\evo{\Delta h^2_s}_i$) is the spatial (energy)
variance of DLWF $i$, $0 \leq \gamma \leq 1$,
$C = \SI{1}{\bohr^2/\electronvolt^2}$, and $\si{\bohr}$ is the Bohr radius.

\begin{figure}[ht]
    \centering
    \includegraphics[width=0.7\textwidth]{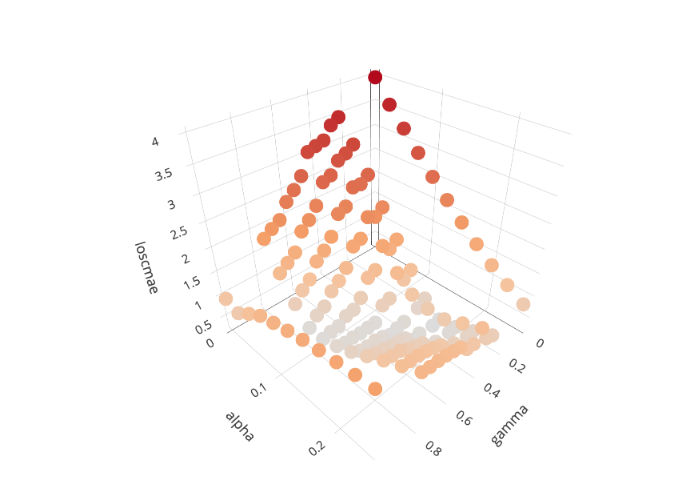}
    \caption{The mean absolute error (MAE, eV) in the band gap as a function of
             $\gamma$ and $\alpha$. Orange points have larger MAE, while grey
             points have smaller MAE.
             }\label{fig:gamma-sweep}
\end{figure}

As recommended by one of the referees, we investigated the effect of varying
$\gamma$, which controls the relative importance of spatial and energy
localization, on a subset of materials in the dataset:\ GaSb, Ge, InAs, InN,
InSb, NaF, Si, SiC. $\gamma$ was varied between 0, the maximally localized
Wannier functions (MLWFs) of \citet{marzari_maximally_1997}, and 1, yielding
pure Fourier transforms of the disentangled Bloch orbitals
\cite{souza_maximally_2001} (up to phases at each $\bk$-point). At each value
of $\gamma$, the screening parameter $\alpha$ was scanned from
\SIrange{0.000}{0.250}{\bohr^{-1}}. Observe from Fig.\ \ref{fig:gamma-sweep}
that the optimal $\alpha$ required increases as $\gamma$ decreases; in other
words, more screening is required for accurate band gaps when the sLOSC
correction is produced by more spatially localized orbitals. Note that setting
$\gamma > 0.50$ resulted in DLWFs that were not well localized within the
$6 \times 6 \times 6$ Born--von Karman supercell; they would require an even
larger supercell to give accurate answers, which is computationally expensive.

Therefore, we chose six combinations of $\gamma \leq 0.50$ and $\alpha$ with
small MAE in the bulk subset and obtained the mean absolute percentage error
(MAPE) of the molecular systems (those in Table \ref{tab:mol-data} below) at
those values. There is not an appreciable change compared to the $\gamma$ and
$\alpha$ in the main text, which is approximated by the $(0.50, 0.150)$ data
point. Thus, we are satisfied that the value of $\gamma$ in the main text is
good for the sLOSC method.

\begin{longtable}{@{}ccc@{}}
  \caption{The MAPE in the fundamental gap of the molecular systems as $\gamma$
           and $\alpha$ are varied.
           }\label{tab:mol-gamma}\\
  \hline\hline
  $\gamma$ & $\alpha$ (\si{\bohr^{-1}}) & MAPE (\%) \\ 
  \colrule
  0.25 & 0.175 & 46.08 \\ 
  0.30 & 0.175 & 47.59 \\ 
  0.40 & 0.175 & 48.58 \\ 
  0.45 & 0.150 & 43.41 \\ 
  0.45 & 0.175 & 49.58 \\ 
  0.50 & 0.150 & 43.68 \\ 
  \hline\hline
\end{longtable}

\section{sLOSC data} \label{sec:data}
sLOSC is evaluated on a modified version of the SC/40 test set found in
\cite{heyd_energy_2005}, with modifications as follows: Systems for which no
experimental lattice constant could be found (BSb, CaTe, SrTe) were excluded;
and the large-gap insulators LiF, LiCl, NaF, NaCl, Ar, Ne were added, for a
total of 43 systems. Computations were carried out in the
\texttt{Quantum ESPRESSO} code suite
\cite{giannozzi_quantum_2009,giannozzi_advanced_2017}, to which sLOSC is
implemented as an add-on package, and a version of the \texttt{wannier90} code 
\cite{mostofi_wannier90_2008, mostofi_updated_2014, pizzi_wannier90_2020}
modified for the computation of DLWFs \cite{mahler_wannier_2022}.

For every system in which \cite{heyd_energy_2005} lists an experimental lattice
constant, we use it; the other experimental lattice constants are sourced from
\cite{wyckoff_crystal_1973}. Experimental band gaps are also sourced from
\cite{heyd_energy_2005} whenever they are available; for the rest of the
systems, the experimental gaps are sourced from \cite{tran_accurate_2009}
(LiF, LiCl, Ne, Ar); \cite{poole_electronic_1975} (NaF);
\cite{roessler_electronic_1968} (NaCl); and \cite{kaneko_new_1988}
(CaS, CaSe, SrS, SrSe).

\begin{longtable}{@{}lcrrrrrr@{}}
    \caption{The data presented in Fig.\ 1 of the main text. The second column
             gives the Strukturbericht designation for each system; note that
             B4 (wurtzite) structures have two lattice constants.}\\
    \label{tab:sc40-data} \\
    \hline\hline
    \multicolumn{3}{c}{System parameters} & \multicolumn{3}{c}{Band gaps (\si{\electronvolt})} & \multicolumn{2}{c}{Energy (\si{\electronvolt})} \\
    \cmidrule(lr){1-3} \cmidrule(lr){4-6} \cmidrule(lr){7-8}
    \multicolumn{1}{c}{Formula} & \multicolumn{1}{c}{Lattice} & \multicolumn{1}{c}{Constant (\si{\angstrom})} & \multicolumn{1}{c}{DFA} & \multicolumn{1}{c}{sLOSC} & \multicolumn{1}{c}{Exp.} & \multicolumn{1}{c}{$E_{\text{DFA}}$} & \multicolumn{1}{c}{$\Delta E_{\text{sLOSC}}$} \\
    \colrule
    \endfirsthead
    \hline\hline
    \multicolumn{3}{c}{System parameters} & \multicolumn{3}{c}{Band gaps (\si{\electronvolt})} & \multicolumn{2}{c}{Energy (\si{\electronvolt})} \\
    \cmidrule(lr){1-3} \cmidrule(lr){4-6} \cmidrule(lr){7-8}
    \multicolumn{1}{c}{Formula} & \multicolumn{1}{c}{Lattice} & \multicolumn{1}{c}{Constant (\si{\angstrom})} & \multicolumn{1}{c}{DFA} & \multicolumn{1}{c}{sLOSC} & \multicolumn{1}{c}{Exp.} & \multicolumn{1}{c}{$E_{\text{DFA}}$} & \multicolumn{1}{c}{$\Delta E_{\text{sLOSC}}$} \\
    \colrule
    \endhead
  AlAs & B3 & 5.661 & 1.440 & 2.828 & 2.23 & $-$2950.23 & $6.550 \times 10^{-3}$ \\ 
  AlN & B4 & (a) 3.111 & 4.306 & 7.296 & 6.13 & $-$696.28 & $4.290 \times 10^{-3}$ \\ 
      &    & (c) 4.981 \\
  AlP & B3 & 5.463 & 1.580 & 2.968 & 2.51 & $-$252.76 & $5.886 \times 10^{-3}$ \\ 
  AlSb & B3 & 6.136 & 1.221 & 2.032 & 1.68 & $-$2538.75 & $1.074 \times 10^{-2}$ \\ 
  Ar & A1 & 5.256 & 8.703 & 12.017 & 14.20 & $-$614.70 & $2.626 \times 10^{-4}$ \\ 
  BaS & B1 & 6.389 & 2.148 & 3.903 & 3.88 & $-$1068.90 & $1.382 \times 10^{-2}$ \\ 
  BAs & B3 & 4.777 & 1.294 & 2.490 & 1.46 & $-$2966.27 & $2.450 \times 10^{-3}$ \\ 
  BaSe & B1 & 6.595 & 1.922 & 3.437 & 3.58 & $-$4067.49 & $1.465 \times 10^{-2}$ \\ 
  BaTe & B1 & 7.007 & 1.561 & 2.736 & 3.08 & $-$3621.33 & $1.290 \times 10^{-2}$ \\ 
  BN & B3 & 3.616 & 4.533 & 6.012 & 6.22 & $-$364.71 & $1.429 \times 10^{-4}$ \\ 
  BP & B3 & 4.538 & 1.348 & 2.053 & 2.40 & $-$269.31 & $8.857 \times 10^{-4}$ \\ 
  C & A4 & 3.567 & 4.206 & 4.712 & 5.48 & $-$327.57 & $8.299 \times 10^{-5}$ \\ 
  CaS & B1 & 5.689 & 2.358 & 5.139 & 4.43 & $-$1358.08 & $7.687 \times 10^{-3}$ \\ 
  CaSe & B1 & 5.916 & 2.035 & 4.567 & 3.85 & $-$4356.48 & $1.220 \times 10^{-2}$ \\ 
  CdS & B3 & 5.818 & 1.163 & 2.291 & 2.55 & $-$5072.67 & $1.416 \times 10^{-2}$ \\ 
  CdSe & B3 & 6.052 & 0.639 & 1.571 & 1.90 & $-$8071.28 & $1.732 \times 10^{-2}$ \\ 
  CdTe & B3 & 6.480 & 0.773 & 1.414 & 1.92 & $-$7625.31 & $1.835 \times 10^{-2}$ \\ 
  GaAs & B3 & 5.648 & 0.538 & 1.224 & 1.52 & $-$4967.79 & $8.015 \times 10^{-3}$ \\ 
  GaN & B4 & (a) 3.189 & 1.941 & 4.177 & 3.50 & $-$4728.22 & $7.120 \times 10^{-3}$ \\ 
         &    & (c) 5.185 \\
  $\beta$-GaN & B3 & 4.523 & 1.689 & 4.007 & 3.30 & $-$2364.11 & $3.380 \times 10^{-3}$ \\ 
  GaP & B3 & 5.451 & 1.658 & 2.664 & 2.35 & $-$2270.19 & $9.008 \times 10^{-3}$ \\ 
  GaSb & B3 & 6.096 & 0.116 & 0.560 & 0.73 & $-$4556.57 & $7.704 \times 10^{-3}$ \\ 
  Ge & A4 & 5.658 & 0.031 & 0.222 & 0.74 & $-$4861.09 & $3.093 \times 10^{-3}$ \\ 
  InAs & B3 & 6.058 & 0.000 & 0.202 & 0.41 & $-$4694.91 & $1.463 \times 10^{-2}$ \\ 
  InN & B4 & (a) 3.537 & 0.000 & 1.440 & 0.69 & $-$4180.63 & $5.138 \times 10^{-2}$ \\ 
      &    & (c) 5.704 \\
  InP & B3 & 5.869 & 0.689 & 1.402 & 1.42 & $-$1997.15 & $9.919 \times 10^{-3}$ \\ 
  InSb & B3 & 6.479 & 0.000 & 0.260 & 0.23 & $-$4283.83 & $1.943 \times 10^{-2}$ \\ 
  LiCl & B1 & 5.130 & 6.328 & 8.877 & 9.40 & $-$645.92 & $9.402 \times 10^{-4}$ \\ 
  LiF & B1 & 4.017 & 9.186 & 14.387 & 14.20 & $-$870.52 & $1.918 \times 10^{-4}$ \\ 
  MgO & B1 & 4.207 & 4.803 & 8.623 & 7.22 & $-$2069.45 & $6.925 \times 10^{-4}$ \\ 
  MgS & B3 & 5.622 & 3.558 & 5.273 & 5.40 & $-$1919.41 & $4.867 \times 10^{-3}$ \\ 
  MgSe & B1 & 5.400 & 1.863 & 3.235 & 2.47 & $-$4917.70 & $5.720 \times 10^{-3}$ \\ 
  MgTe & B3 & 6.420 & 2.499 & 3.773 & 3.60 & $-$4471.36 & $6.453 \times 10^{-3}$ \\ 
  NaCl & B1 & 5.641 & 5.107 & 7.419 & 8.97 & $-$1693.81 & $1.595 \times 10^{-3}$ \\ 
  NaF & B1 & 4.620 & 6.389 & 11.610 & 11.50 & $-$1917.95 & $5.157 \times 10^{-4}$ \\ 
  Ne & A1 & 4.429 & 11.617 & 18.101 & 21.70 & $-$907.70 & $6.803 \times 10^{-5}$ \\ 
  Si & A4 & 5.430 & 0.709 & 1.572 & 1.17 & $-$230.28 & $6.914 \times 10^{-3}$ \\ 
  SiC & B3 & 4.358 & 1.363 & 1.962 & 2.42 & $-$279.46 & $1.000 \times 10^{-3}$ \\ 
  SrS & B1 & 5.990 & 2.455 & 4.434 & 4.32 & $-$1236.19 & $6.467 \times 10^{-3}$ \\ 
  SrSe & B1 & 6.234 & 2.192 & 3.953 & 3.81 & $-$4234.69 & $7.301 \times 10^{-3}$ \\ 
  ZnS & B3 & 5.409 & 2.101 & 3.300 & 3.66 & $-$5990.28 & $6.740 \times 10^{-3}$ \\ 
  ZnSe & B3 & 5.668 & 1.289 & 2.301 & 2.70 & $-$8988.74 & $9.112 \times 10^{-3}$ \\ 
  ZnTe & B3 & 6.089 & 1.282 & 1.978 & 2.38 & $-$8542.59 & $1.592 \times 10^{-2}$ \\ 
\hline\hline
\end{longtable}

We additionally compute sLOSC results on a small set of molecular systems
using the in-house QM$^4$D code. We use the PBE exchange-correlation functional,
as in the bulk case, and the 6-311++G(3df,3pd) basis set, with the Dunning
augmented correlation-consistent triple-zeta (aug-cc-pVTZ) basis used
for density fitting.

Instead of experimental values, we use CCSD(T) computations for the reference
fundamental gaps. These are computed under both the aug-cc-pVTZ and aug-cc-pVQZ
(Dunning augmented correlation-consistent triple- and quadruple-zeta) basis sets
and extrapolated using the formula from Eq.\ (44) of \cite{lin_long-range_2012} as
\begin{equation}
    \epsilon_\infty 
        = \frac{\epsilon_{\text{TZ}}\times 3^3-\epsilon_{\text{QZ}}\times 4^3}
            {3^3 - 4^3},
\end{equation}
where $\epsilon_{\text{nZ}}$ is the aug-cc-pVnZ eigenvalue.

\begin{longtable}{@{}lrrrrr@{}}
    \caption{The molecular sLOSC data.}\\
    \label{tab:mol-data} \\
    \hline\hline
    & \multicolumn{3}{c}{Band gaps (\si{\electronvolt})} & \multicolumn{2}{c}{Energy (\si{\electronvolt})} \\
    \cmidrule(lr){2-4} \cmidrule(lr){5-6}
    \multicolumn{1}{c}{Formula} & \multicolumn{1}{c}{DFA} & \multicolumn{1}{c}{sLOSC} & \multicolumn{1}{c}{Exp.} & \multicolumn{1}{c}{$E_{\text{DFA}}$} & \multicolumn{1}{c}{$\Delta E_{\text{sLOSC}}$} \\
    \colrule
    \endfirsthead
    \hline\hline
    & \multicolumn{3}{c}{Band gaps (\si{\electronvolt})} & \multicolumn{2}{c}{Energy (\si{\electronvolt})} \\
    \cmidrule(lr){2-4} \cmidrule(lr){5-6}
    \multicolumn{1}{c}{Formula} & \multicolumn{1}{c}{DFA} & \multicolumn{1}{c}{sLOSC} & \multicolumn{1}{c}{Exp.} & \multicolumn{1}{c}{$E_{\text{DFA}}$} & \multicolumn{1}{c}{$\Delta E_{\text{sLOSC}}$} \\
    \colrule
    \endhead
  \ce{SH} & 0.430 & 3.676 & 8.05 & $-$10845.99 & $2.346 \times 10^{-4}$ \\ 
  \ce{O2} & 2.306 & 6.959 & 12.51 & $-$4088.33 & $2.072 \times 10^{-5}$ \\ 
  \ce{S2} & 1.292 & 2.497 & 7.92 & $-$21662.11 & $2.906 \times 10^{-4}$ \\ 
  \ce{Cl2} & 2.740 & 4.395 & 10.61 & $-$25035.22 & $1.035 \times 10^{-4}$ \\ 
  \ce{CF2} & 3.626 & 7.544 & 12.68 & $-$6464.52 & $5.676 \times 10^{-4}$ \\ 
  \ce{CH2} & 2.413 & 6.845 & 9.57 & $-$1064.09 & $9.100 \times 10^{-4}$ \\ 
  \ce{CH3} & 2.472 & 6.531 & 9.84 & $-$1082.70 & $1.029 \times 10^{-3}$ \\ 
  \ce{CN} & 1.709 & 6.125 & 10.29 & $-$2520.88 & $2.064 \times 10^{-4}$ \\ 
  \ce{CHO} & 1.693 & 4.568 & 9.44 & $-$3095.93 & $1.407 \times 10^{-3}$ \\ 
  \ce{CH3O} & 0.723 & 6.102 & 9.51 & $-$3128.24 & $2.093 \times 10^{-3}$ \\ 
  \ce{CH2S} & 1.891 & 4.067 & 9.16 & $-$11898.67 & $9.618 \times 10^{-4}$ \\ 
  \ce{CH2SH} & 1.567 & 3.743 & 5.91 & $-$11914.33 & $9.567 \times 10^{-4}$ \\ 
  \ce{OH} & 0.916 & 8.599 & 11.24 & $-$2059.42 & $1.223 \times 10^{-4}$ \\ 
  \ce{NH} & 3.597 & 9.751 & 13.15 & $-$1501.30 & $3.693 \times 10^{-4}$ \\ 
  \ce{NH2} & 2.738 & 8.527 & 11.37 & $-$1519.24 & $1.025 \times 10^{-3}$ \\ 
  \ce{SiH2} & 1.813 & 3.671 & 8.48 & $-$7903.86 & $4.788 \times 10^{-4}$ \\ 
  \ce{SiH3} & 1.828 & 3.924 & 7.93 & $-$7920.69 & $1.044 \times 10^{-3}$ \\ 
  \ce{PH} & 2.213 & 4.728 & 9.16 & $-$9298.83 & $1.938 \times 10^{-4}$ \\ 
  \ce{PH2} & 1.844 & 4.349 & 8.56 & $-$9315.89 & $7.920 \times 10^{-4}$ \\ 
  \hline\hline
\end{longtable}

\begin{figure}[!htp]
\centering
\includegraphics[width=0.8\linewidth]{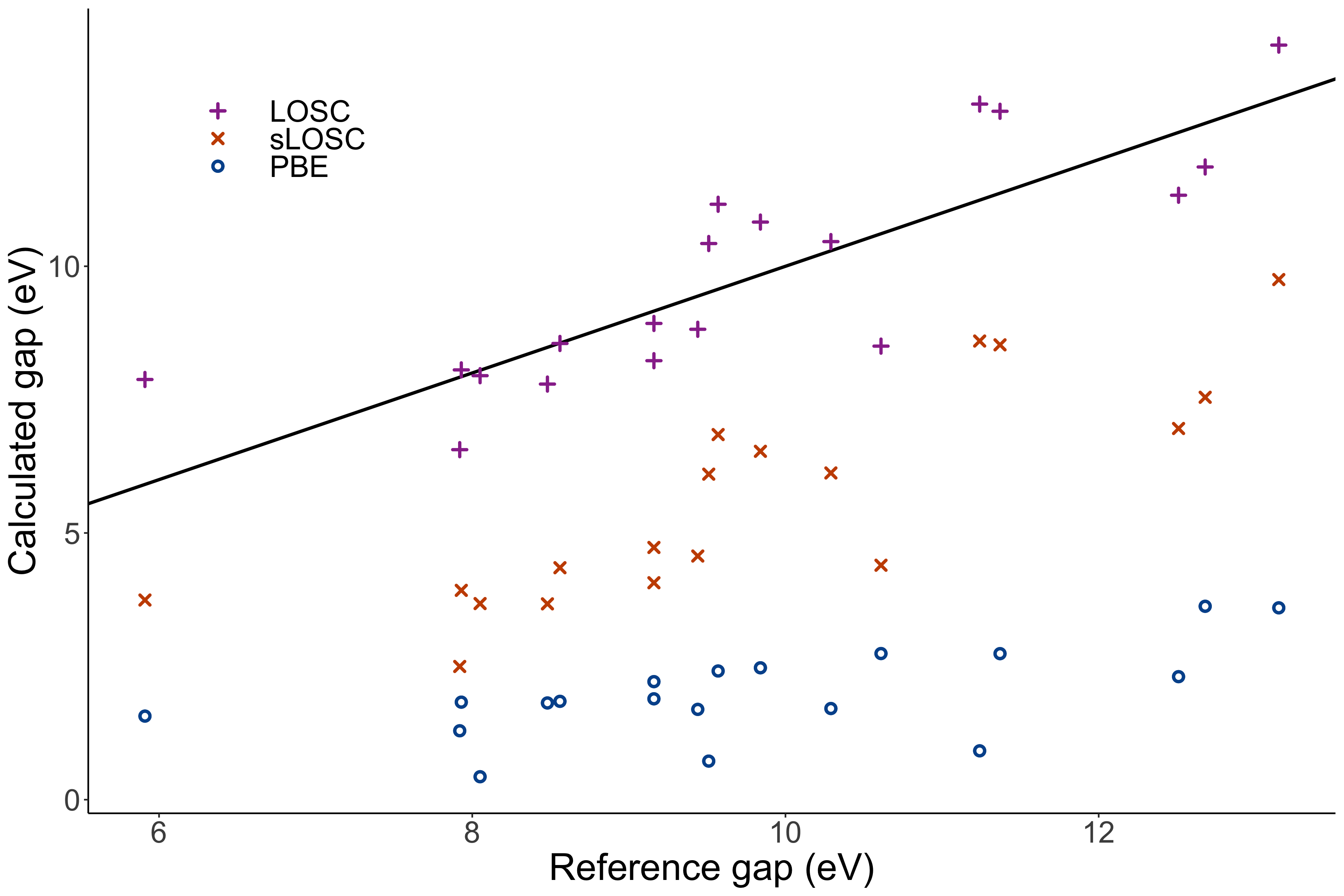}
\caption{Comparison of experimental fundamental gaps of molecular systems with
         those calculated by PBE (\texttt{o}), sLOSC ($\boldsymbol{\times}$),
         and unscreened LOSC (\texttt{+}).} 
\label{fig:scatter-mol}
\end{figure}

We plot the data from Table \ref{tab:mol-data} in Fig.\ \ref{fig:scatter-mol},
together with PBE and unscreened LOSC results, in analogy with Fig.\ 1 in the
main text. We can see again that the best gaps are uniformly those computed
without screening, in agrement with Fig. \ref{fig:screensweep} and as expected
for small molecules with open boundary conditions; however, sLOSC still improves
the accuracy of the gap compared to the PBE calculation.

\clearpage
\bibliography{supplemental}
\bibliographystyle{apsrev4-2}